\documentclass[fleqn,usenatbib]{mnras}

\usepackage{newtxtext,newtxmath}
\usepackage[T1]{fontenc}
\usepackage{ulem}

\DeclareRobustCommand{\VAN}[3]{#2}
\let\VANthebibliography\thebibliography
\def\thebibliography{\DeclareRobustCommand{\VAN}[3]{##3}\VANthebibliography}

\usepackage{graphicx}	% Including figure files
\usepackage{amsmath}	% Advanced maths commands
\usepackage{rotating}
\usepackage{gensymb}

\title[Optical/nIR counterparts of GW190814]{A search for optical and near-infrared counterparts of the compact binary merger GW190814}

\author[A. L. Thakur et al.]{A.~L. Thakur$^{1,2}$\thanks{E-mail: aishwarya.thakur@inaf.it},
S. Dichiara$^{3,4}$, 
E. Troja$^{3,4}$, 
E.~A. Chase$^{5,6,7,8}$, 
R. S\'{a}nchez-Ram\'{i}rez$^{1}$, 
\newauthor{L. Piro$^{1}$,
C.~L. Fryer$^{5,9,10,11,12,13}$, 
N. R. Butler$^{14}$,
A. M. Watson $^{15}$,
R.~T. Wollaeger$^{5,10}$,}
\newauthor{E. Ambrosi$^{16}$, 
J. Becerra González$^{17,18}$,
R. L. Becerra $^{19}$,
G. Bruni$^{1}$, 
S. B. Cenko $^{4, 20}$, } 
\newauthor{G. Cusumano$^{16}$,
A. D'Aì$^{16}$, 
J. Durbak $^{3}$, 
C.~J. Fontes$^{5,6}$, 
P. Gatkine$^{3}$,
A.~L. Hungerford$^{5,6,9}$,}
\newauthor{O. Korobkin$^{5,9,10}$, 
A. S. Kutyrev$^{4}$,
W. H. Lee$^{15}$,
S. Lotti$^{1}$,
G. Minervini$^{1}$, 
G. Novara$^{21,22}$,} 
\newauthor{V. La Parola$^{16}$
M. Pereyra$^{23}$,
R. Ricci$^{24,25}$,
A. Tiengo$^{21,22,26}$, 
S. Veilleux$^{3}$}
\\
$^{1}$INAF-Istituto di Astrofisica e Planetologia Spaziali, via Fosso del Cavaliere, 100, I-00133 Rome RM, Italy\\
$^{2}$Dipartimento di Fisica, Universit\`a degli Studi di Roma "Tor Vergata", Via della Ricerca Scientifica 1, 00133 Roma RM, Italy\\
$^{3}$Department of Astronomy, University of Maryland, College Park, MD 20742-4111, USA\\
$^{4}$Astrophysics Science Division, NASA Goddard Space Flight Center, 8800 Greenbelt Rd, Greenbelt, MD 20771, USA\\
$^{5}$Center for Theoretical Astrophysics, Los Alamos National Laboratory, Los Alamos, NM, 87545, USA\\
$^{6}$Computational Physics Division, Los Alamos National Laboratory, Los Alamos, NM, 87545, USA\\
$^{7}$Center for Interdisciplinary Exploration and Research in Astrophysics (CIERA), Northwestern University, Evanston, IL, 60201, USA\\
$^{8}$Department of Physics and Astronomy, Northwestern University, Evanston, IL 60208, USA\\
$^{9}$Joint Institute for Nuclear Astrophysics - Center for the Evolution of the Elements, USA\\
$^{10}$Computer, Computational, and Statistical Sciences Division, Los Alamos National Laboratory, Los Alamos, NM, 87545, USA\\
$^{11}$Steward Observatory, The University of Arizona, Tucson, AZ 85721, USA\\
$^{12}$Department of Physics and Astronomy, The University of New Mexico, Albuquerque, NM 87131, USA\\
$^{13}$The George Washington University, Washington, DC 20052, USA\\
$^{14}$School of Earth and Space Exploration, Arizona State University, Tempe, AZ 85287, USA\\
$^{15}$Instituto de Astronom{\'\i}a, Universidad Nacional Autónoma de M\'exico, Apartado Postal 70-264, 04510 M\'exico, CDMX, Mexico\\
$^{16}$INAF-Istituto di Astrofisica Spaziale e Fisica Cosmica, Via Ugo la Malfa, 153, I-90146 Palermo PA, Italy\\
$^{17}$Universidad de La Laguna, Dpto. Astrofísica, E-38206 La Laguna, Tenerife, Spain\\
$^{18}$Instituto de Astrofísica de Canarias, E-38200 La Laguna, Tenerife, Spain\\
$^{19}$Instituto de Ciencias Nucleares, Universidad Nacional Autónoma de México, Apartado Postal 70-543, 04510 CDMX, México\\
$^{20}$Joint Space-Science Institute, University of Maryland, College Park, Maryland 20742, USA\\
$^{21}$Scuola Universitaria Superiore IUSS Pavia, Piazza della Vittoria 15, I-27100, Pavia, Italy\\
$^{22}$INAF, Istituto di Astrofisica Spaziale e Fisica Cosmica Milano, via A.,Corti 12, I-20133 Milano, Italy\\
$^{23}$CONACYT, Instituto de Astronomía, Universidad Nacional Autónoma de México, 22860 Ensenada, BC, Mex\\
$^{24}$INAF-Istituto di Radioastronomia, Via Gobetti 101, I-40129, Bologna, Italy\\
$^{25}$Istituto Nazionale di Ricerca Metrologica (INRiM) - Strada delle Cacce 91, I-10135 Torino, Italy\\
$^{26}$INFN, Sezione di Pavia, via A. Bassi 6, I-27100 Pavia, Italy\\
}
\pubyear{2020}

\begin{document}
%\label{firstpage}
%\pagerange{\pageref{firstpage}--\pageref{lastpage}}
\maketitle

%\nocite{*}

% Abstract of the paper
\begin{abstract}
We report on our observing campaign of the compact binary merger GW190814, detected by the Advanced LIGO and Advanced Virgo detectors on August 14th, 2019. This signal has the best localisation of any observed gravitational wave (GW) source, with a 90\% probability area of 18.5~deg$^{2}$, and an estimated distance of $\approx$240~Mpc.
We obtained wide-field observations with the Deca-Degree Optical Transient Imager (DDOTI) covering 88\% of the probability area down to a limiting magnitude of $w$ = 19.9 AB. Nearby galaxies within the high probability region were targeted with the Lowell Discovery Telescope (LDT),
whereas  promising candidate counterparts were characterized through multi-colour photometry with the Reionization and Transients InfraRed (RATIR) and spectroscopy with the Gran Telescopio de Canarias (GTC). We use our optical and near-infrared limits in conjunction with the upper limits obtained by the community to constrain the possible electromagnetic counterparts associated with the merger. 
A gamma-ray burst seen along its jet's axis is disfavoured by the multi-wavelength dataset, whereas the presence of a burst seen at larger viewing angles is not well constrained. 
Although our observations are not sensitive to a kilonova similar to AT2017gfo, we can rule out high-mass (> 0.1 M$_{\odot}$) fast-moving (mean velocity $\geq$ 0.3$c$) wind ejecta for a possible kilonova associated with this merger.

\end{abstract}

% Select between one and six entries from the list of approved keywords.
% Don't make up new ones.

%%%%%%%%%%%%%%%%%%%%%%%%%%%%%%%%%%%%%%%%%%%%%%%%%%

%%%%%%%%%%%%%%%%% BODY OF PAPER %%%%%%%%%%%%%%%%%%

\begin{figure*}
\centering
\includegraphics[trim=15 15 30 5, width=\linewidth]{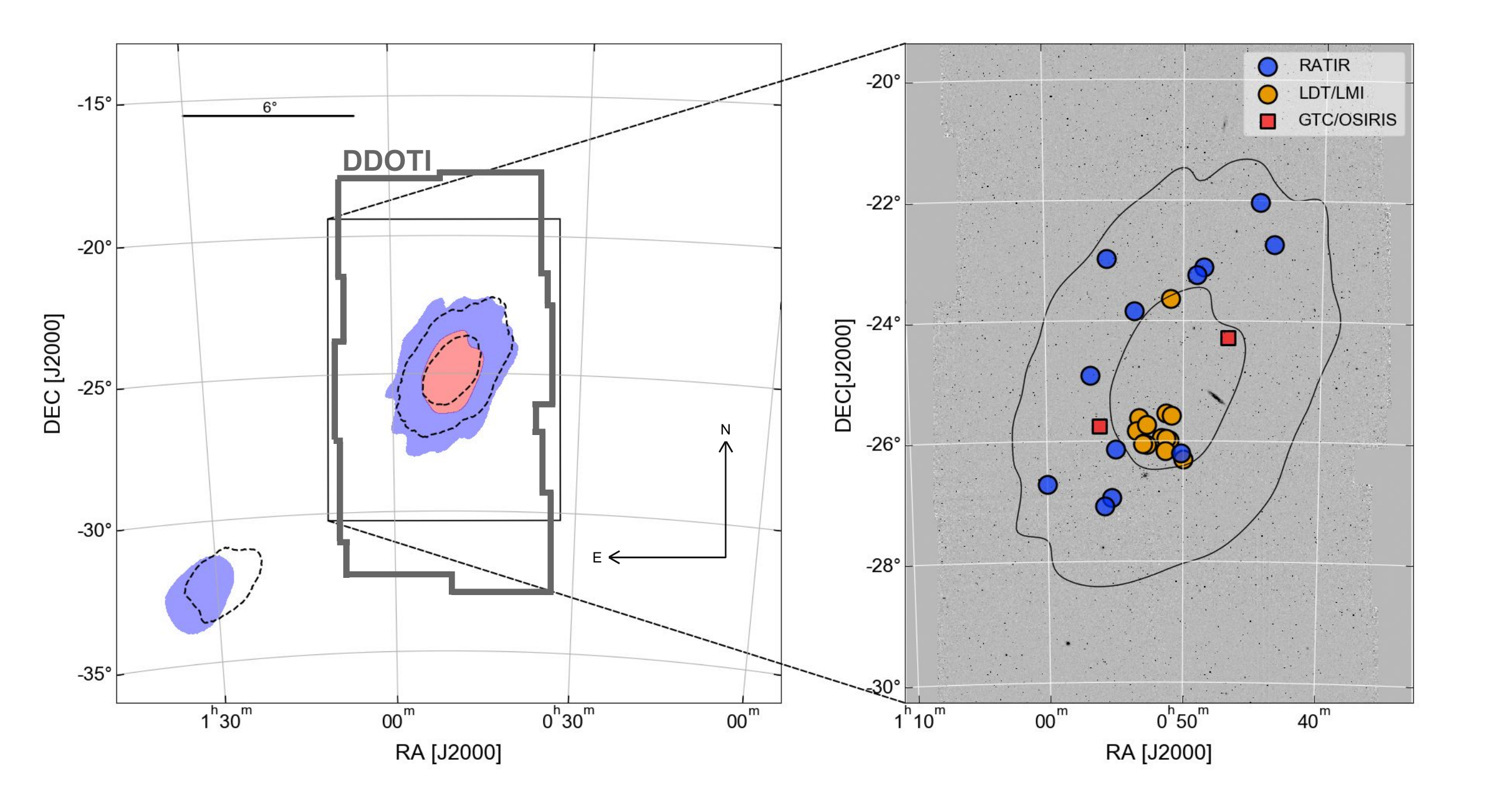}
\caption{{\it Left panel}: LALInference skymap for GW190814. The
colored contours show the 50\% (red) and the 90\% (purple) probability regions. 
The dashed line shows the refined localisation from \citet{Abbott_2020_GW190814bv}. 
The area observed by DDOTI is also shown (thick line). 
{\it Right panel}:
DDOTI image of the field of GW190814 . The targets observed with RATIR (blue), LDT/LMI (orange) and GTC/OSIRIS (red) are shown. The 50\% and 90\% localisation areas of the LALInference skymap are superimposed on the image for reference. 
}
\label{fig:ddoti}
\end{figure*}

\section{Introduction}\label{sec:int}

The era of gravitational wave (GW) astronomy started with the discovery of GW150914 \citep{PhysRevLett.116.061102}, which provided the first direct GW detection from a compact binary coalescence (CBC). The signal was generated by the merger of a binary black hole (BBH) to form a single black hole. 
Two years later, the detection of a GW signal from a binary neutron star (BNS) merger \citep[GW170817;][]{PhysRevLett.119.161101} led to another fundamental breakthrough. The detection of the short gamma-ray burst (sGRB) and kilonova  \citep[GRB 170817A and AT2017gfo, respectively.][]{Abbott_2017} associated with GW170817 provided the first observations of the electromagnetic (EM) counterparts of a GW source. 
During the first (O1) and second (O2) observing runs of the Advanced LIGO \citep{2015CQGra..32g4001L} and Advanced Virgo \citep{2015CQGra..32b4001A} detectors, 11 confirmed detections of GW signals from CBCs were reported, of which, ten were BBH mergers and one was a BNS merger \citep{PhysRevX.9.031040}. The third observing run (O3) started on  April 1st, 2019 and was suspended on March 27th, 2020. At the time of writing, O3 data have produced four confirmed detections \citep[GW190412, GW190425, GW190521, GW190814;][respectively]{2020arXiv200408342T, Abbott_2020, PhysRevLett.125.101102,Abbott_2020_GW190814bv} and 54 unretracted candidate signals\footnote{https://gracedb.ligo.org/superevents/public/O3/.}

GW190814 was observed on August 14th, 2019 at 21:10:39 UTC. The initial analysis with BAYESTAR \citep{PhysRevD.93.024013} used data from the LIGO-Livingston and Virgo detectors, which resulted in a $90\%$ localisation area of 772 deg$^2$ and mean distance of 276 Mpc with a standard deviation of 56 Mpc. Data from the LIGO-Hanford detector were later added to the analysis and resulted in an updated 90$\%$ localisation value of 38 deg$^2$.
The candidate was initially assigned a high probability Mass Gap classification based on preliminary analysis \citep[> 99\%;][]{2019GCN.25324....1L}. A Mass-Gap classification suggests that the mass of at least one of the binary components is between 3 and 5 M$_{\sun}$. Upon further analysis with LALInference \citep{PhysRevD.91.042003}, the signal classification was updated to a neutron star - black hole (NSBH) merger \citep{2019GCN.25333....1L}. The mass of the lighter object was reported to be $<$ 3 M$_{\odot}$ and the mass of the heavier object to be $>$ 5 M$_{\odot}$. The NSBH classification was based on the assumption that the heavier object is a black hole and the lighter object is a neutron star. The localisation and distance estimate were also refined during this analysis with an updated distance estimate of 267 ($\pm$ 52) Mpc and a 90$\%$ (50$\%$) localisation value of 23 (5) deg$^{2}$. 
These values were only slightly modified in the final analysis, 
presented in \citet{Abbott_2020_GW190814bv}. The median distance reported for this merger is 241$_{-45}^{+41}$~Mpc, and its localisation further improved to 18.5~deg$^{2} $(90\%). 
The heavier object is a BH with mass 23.2$_{-1.0}^{+1.1}$ M$_\odot$, whereas the lighter object, with a mass of 2.6$_{-0.09}^{+0.08}$ M$_\odot$, is not unambiguously classified. 

The low false-alarm rate (FAR) and its preliminary association to an NSBH merger make GW190814 an event of considerable interest, although the mass of the lighter object does not rule out a BBH. The localisation area for this event is the best for any GW signal so far, and allowed for extensive follow-up observations to search for possible electromagnetic (EM) counterparts \citep[e.g.,][]{2020arXiv200201950A,andreoni2019growth,Dobie19}. 

The detection of any EM counterpart helps improve the localisation of a GW signal while simultaneously providing information on the physics of the merger and its environment \citep{PhysRevLett.119.161101}. 
Whereas the EM signatures of a BBH merger \citep[e.g.,][]{Graham20, PhysRevLett.125.101102} are uncertain, the possible counterparts of an NSBH merger should be in many ways similar to the EM signals associated to BNS mergers. A short duration gamma-ray burst (sGRB), produced by a relativistic jet launched from the merger remnant, may be visible soon after the merger \citep[e.g., GRB 170817A;][]{Abbott_2017}. The interaction of this relativistic jet with the circumburst environment produces afterglow emission, observable across the EM spectrum \citep[e.g., ][]{Troja2017, Hallinan1579}. 

Dynamical ejecta and sub-relativistic wind outflows produce a distinctive EM signal known as a kilonova \citep[KN,][]{Li_1998, Metzger2019}. The composition of heavy elements synthesized via the r-process determines the emergent spectrum. High-opacity lanthanides from neutron rich material, (electron fraction $Y_e\lesssim0.3$) give rise to a red component, while material with higher electron fraction produces a blue component \citep[][]{Barnes2013,Kasen2013ApJ,kasen15,kasen17,Tanaka2017,2018MNRAS.478.3298W}. Tidal ejecta are dominated by neutron rich material while disk winds exhibit a broad range of  $Y_e$, and are thus able to support both a blue and a red component \citep{kasen17,2019PhRvD.100b3008M}. The blue component can be enhanced if the remnant of a BNS merger is a long-lived hyper/supramassive neutron star \citep{Piro2019}. In this case the strong neutrino irradiation would increase the electron fraction  of the polar components of the ejecta, i.e. the wind from the disk and the shock-driven  dynamical ejecta  \citep{sekiguchi16, shibata17, 2019PhRvD.100b3008M}. 
The blue component of  an NSBH kilonova could thus be dimmer in comparison to that from a BNS merger as GW170817 \citep{2020EPJA...56....8B}.

The presence of an EM counterpart in an NSBH is primarily dependent upon the amount of mass left outside the merger remnant, that in turn depends on the equation of state (EOS) of the NS, the mass and spin of the BH, and the orbital characteristics of the encounter \citep{ 2007CQGra..24S.125S,Lee2007,2009PhRvD..79d4024E,2010A&A...514A..66R,Lee2010,2011ApJ...727...95P,2011LRR....14....6S,2014ApJ...780...31T,2015PhRvD..92b4014K,Rosswog_2017,PhysRevD.98.081501,2020PhRvD.101j3002K,2020arXiv200514208F}. These  parameters drive the fraction of the NS material that is tidally disrupted and that remains outside the innermost stable circular orbit (ISCO) of the BH. 
The total mass of ejecta decreases with increasing BH mass, lower spin, and stiffer EOS and drops abruptly to zero once the tidal radius becomes smaller than the BH event horizon. If the NS is tidally disrupted within the ISCO, then no observable signal is expected, contrary to the BNS scenario where a kilonova accompanies mergers of all parameters. 
 
Some numerical studies differentiated from the tidally ejected mass and the disk formed around the BH, a fraction of which produces wind ejecta \citep{2015PhRvD..92b4014K,2020PhRvD.101j3002K,2020arXiv200514208F}. Others published only the total mass not immediately incorporated into the BH \citep{2009PhRvD..79d4024E,2010A&A...514A..66R,2011ApJ...727...95P,2014ApJ...780...31T}. These results typically agree that, if the BH is not spinning, the total mass outside the remnant BH ranges from  roughly 0.1-0.2\,M$_\odot$ for a 4\,M$_\odot$ BH to 0.01\,M$_\odot$ for a 7\,M$_\odot$ BH. Most of this mass ($\gtrsim 60\%$) forms an accretion disk,  dynamical ejecta being about 10-20\% of this total, and wind outflows being typically 10-30\% of the disk mass \citep{2019PhRvD.100b3008M,Metzger2019}. In comparison for GW170817 the mass associated to the red component, i.e. produced by the low $Y_e$ ejecta, was estimated $\approx 0.04$ M$_\odot$ \citep{kasen17}.
In addition to these parameters, the velocity of the various components and other geometrical factors, such as the viewing angle or the shape of dynamical ejecta and the wind, determine the strength  and evolution of the various EM components. 

In this study, we present our search for possible optical and near-infrared counterparts of GW190814. Our campaign encompassed wide-field observations with the Deca-Degree Optical Transient Imager (DDOTI), targeted galaxy observations with the Lowell Discovery Telescope (LDT), photometric and spectroscopic follow up observations of selected candidates with the Reionization and Transients InfraRed (RATIR) and the Gran Telescopio Canarias (GTC) telescopes, respectively. In section \ref{sec:obs}, we describe the observations and data analysis. In section \ref{sec:res} we present the results of our analysis and discuss them in the context of GRB afterglows along with kilonova data in section \ref{sec:dis}. We present our conclusions in section \ref{sec:con}. We note that our calculations are based on the LALInference distance estimate of 267 Mpc which falls within the 90\% confidence interval for the median distance reported in \citet{Abbott_2020_GW190814bv}. Reported photometry values are corrected for the estimated Galactic extinction \citep{Schlafly2011}. Uncertainties are quoted at the 1-$\sigma$ confidence level for each parameter of interest and upper limits are given at a 2-$\sigma$ level, unless stated otherwise. Standard $\Lambda$CDM cosmology \citep{Planck2018} was adopted throughout the paper.

\section{Observations}\label{sec:obs}

Follow-up observations for possible counterparts to a GW signal follow two general strategies: wide-field imaging of the GW localisation area and galaxy-targeted follow-up observations.
In the former case, wide-field imagers are used to perform surveys of the localisation region associated with the signal. 
In the latter case, using a catalogue \citep[see][]{Bilicki_2013, GLADE}, galaxies in the 90$\%$ localisation volume are identified and prioritized based on their probability of hosting the merger. Photometric observations of the selected galaxies are then performed to identify transients possibly associated with the GW candidate signal \citep[See][]{Gomez_2019, 2020arXiv200201950A}. The results of this strategy are affected by the completeness of the galaxy catalogue and the fraction of the total luminosity that is covered. 

After this first step, transient sources showing suitable photometric evolution are identified \citep[For example,][]{andreoni2019growth,2020MNRAS.492.5916W}
and flagged for further observations. This is particularly important to rule out transients like supernovae, which are major contaminants in GW follow-up searches \citep[See][]{Cowperthwaite_2015, Doctor17, andreoni2019growth, 2020arXiv200201950A}.
While spectroscopic follow-up can rapidly determine the distance scale and classify the origin of a transient with a higher degree of certainty, photometric observations can more easily follow a larger number of candidate counterparts. 

We present wide-field observations from DDOTI in section \ref{DDOTIwf}, galaxy targeted observations from LDT in section \ref{LDTGT}, candidate targeted multicolour photometric observations from RATIR in section \ref{RATIRcand} and spectroscopic observations from GTC in section \ref{OSIRISspec}.

\subsection{DDOTI Wide-Field Imaging}\label{DDOTIwf}

The Deca-Degree Optical Transient Imager (DDOTI) employs six 28-cm telescopes with prime focus CCDs mounted on a common equatorial mount. An instantaneous field of view of 69~deg$^{2}$ is obtained by adding together the six field of view of 3.4 $\times$ 3.4~deg on a sky grid of 2 $\times$ 3 \citep{2016SPIE.9910E..0GW}. DDOTI started to observe the main probability region of the updated skymap \citep{2019GCN.25333....1L} on August 15th, 2019 at 7:58 UTC, 10.8 hours after the merger \citep{gcn25352}. The total observed field covers 88\% of the probability in the updated LALInference map (Figure~\ref{fig:ddoti}). This value does not change for the updated skymap in \citet{Abbott_2020_GW190814bv}.

Observations were taken with the airmass ranging between 1.9 and 2.8, a 100\% moon illumination and exposure times between 1020 and 2820 seconds. DDOTI images are unfiltered and photometry measurements are referred to the natural $w$ band.
In our images the number of independent elements inside the 90\% probability area (23 deg$^2$) is $\approx 7.8 \times 10^{6}$, which sets a minimum detection threshold of 6~$\sigma$ for a 99\% confidence level. 
When a reference frame is available, we use this threshold for the analysis of the subtracted image. Otherwise, when comparing our source list to existing catalogues, we follow \citet{2020MNRAS.492.5916W} and adopt a 10-$\sigma$ threshold to filter candidates. We present the results of this analysis in Table \ref{DDOTI}.

\begin{table}
    \centering
    \caption{DDOTI observations}
    \label{DDOTI}
    \begin{tabular}{lccc}
         \hline
         Observation Date & Mid-time (UT) & $w_{max}$ (10-$\sigma$) & $w_{max}$(6-$\sigma$) \\
         \hline
         15 August & 09:55 & 18.5 & 19.0 \\
         16 August & 09:50 & 18.8 & 19.3 \\
         18 August & 09:50 & 18.6 & 19.1 \\
         21 August & 09:30 & 19.9 & - \\
         \hline
    \end{tabular}
    \begin{flushleft}
    \quad
    \footnotesize{
    Column 1: Date of observation;
    Column 2: Midtime of observation; 
    Column 3: 10-$\sigma$ limit; 
    Column 4: Reference image subtracted 6-$\sigma$ limit.}
    \\
    \end{flushleft}
\end{table}

Images from the first night of observations (August 15) were compared with catalogues \citep[USNO-B1 or APASS; ][]{USNO_Monet2003,APASS_Henden2018},
and no potential counterpart was found down to a limiting magnitude of $w_{max}$\,$\sim$18~AB mag (10 $\sigma$; \citealt{2020MNRAS.492.5916W}). 
Additional observations of the field were carried out during the following nights 
(August 16th, 18th, and 21th) using longer exposures (up to 7560 seconds) and reaching deeper field limits of about $w_{max}$=19.9 AB mag (10~$\sigma$). 

This work improves upon previous results from the first night of DDOTI observations \citep{gcn25352,2020MNRAS.492.5916W} as it includes later epochs of observations (August 16th, 18th, and 21th) which were not presented by \citealt{2020MNRAS.492.5916W}. We also used an updated reduction pipeline performing image subtraction and point spread function (PSF)-fitting photometry instead of aperture photometry. These changes improve our sensitivity to transient sources by $\approx 1$ mag with respect to \citet{2020MNRAS.492.5916W}. 
We used the last epoch image (taken on August 21) as template to perform image subtraction on the first night of observations. 
After excluding fast-moving solar system objects and image artifacts, no reliable transient was found in the residual images down to a 6~$\sigma$ limit of $w_{max}$ $\approx$19 AB~mag (this roughly corresponds to a 2-$\sigma$ limits post trial, taking into account the number of independent elements in the field). This limit is $\approx$ 0.7~mag lower for objects in the inner regions of bright galaxies where the bright galaxy's light decreases our sensitivity to point source detection.

The time-gap between the observation of the science and template image is only 6 days. Whereas a rapidly fading kilonova such as AT2017gfo \citep[$\Delta$$m_i$$\approx$2.7 mag between 10 hours and 6 days from the merger][]{Drout17,2017Natur.551...67P,Smartt2017,Troja2017} 
would be detected in our observations, we are not sensitive to slowly evolving transients, such as old supernovae (SNe). 
For example, the bright candidate SN2019mbq \citep[$i$\,$\sim$18.7 AB mag,][]{2019TNSTR1370....1N} is not detected in the subtracted image as its magnitude is nearly constant ($\Delta$$m$$\lesssim$0.1 mag) between our two epochs. The other bright candidate AT2019nqr (desgw-190814d;  $i$\,$\sim$18.3~AB mag) reported by \citet{gcn25373} and later classified as a type II SN \citep{gcn25379}, lies outside the field observed with DDOTI. 
All the other reported candidates are fainter than our limits. Therefore, the lack of candidates in DDOTI observations is consistent with the results reported by other wide-field surveys \citep[e.g., ][]{andreoni2019growth,2020arXiv200201950A}.

\begin{center}
\begin{figure}
\includegraphics[width=\columnwidth]{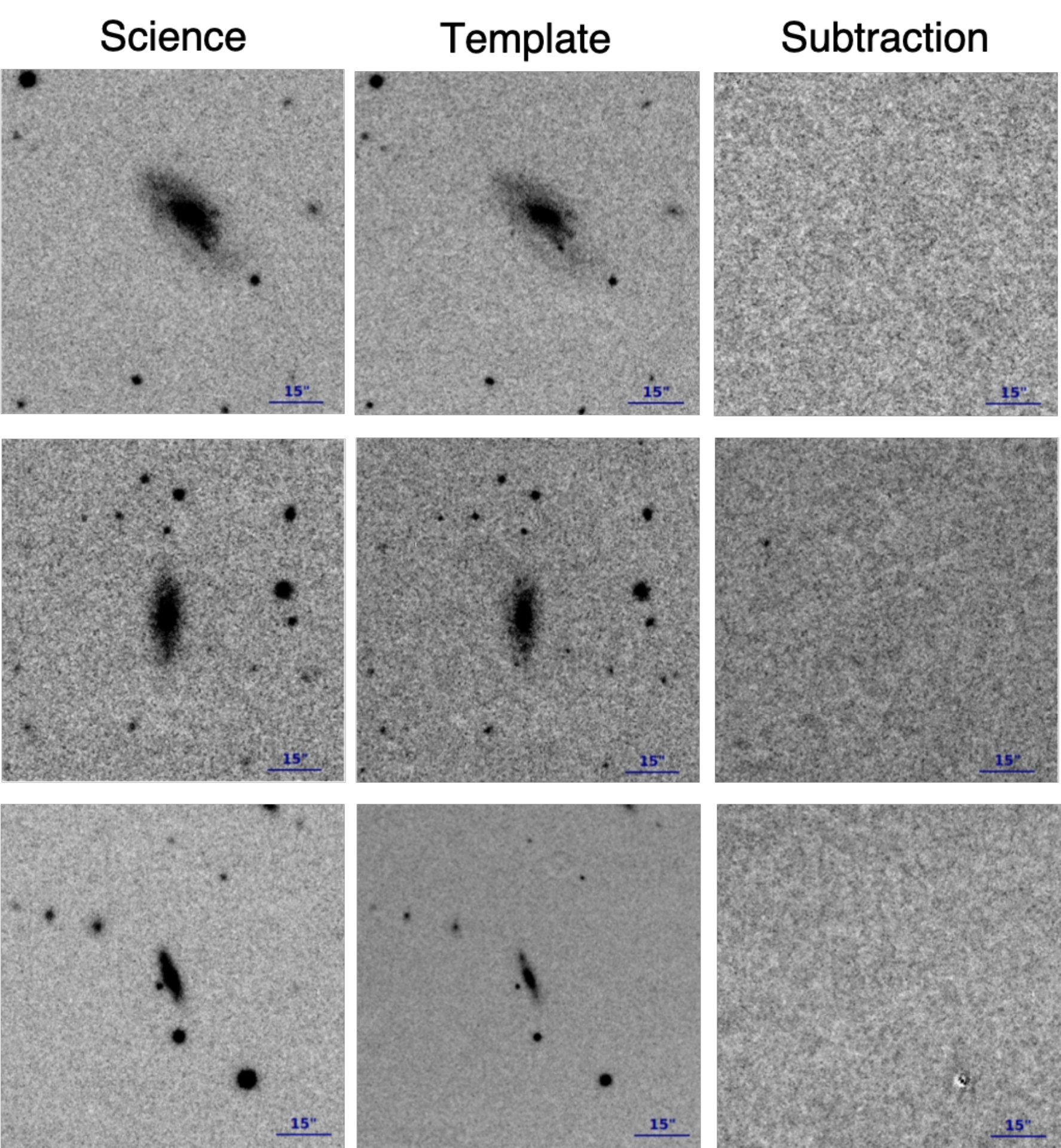}
\caption{Example of galaxies targeted with LDT/LMI: HyperLEDA 776957 (top), HyperLEDA 773149 (middle) and HyperLEDA 777373 (bottom).
 Images were taken at 1.5~d after the merger (science), 3.5~d after the merger (template), and the resulting subtraction is shown in the last column. Images are 3.2$\arcmin$ $\times$ 3.2$\arcmin$ oriented  with North up and East to the left. \label{DCTgalaxy}}
\end{figure}
\end{center}

\begin{figure*}
\begin{center}
\includegraphics[width=0.2\textwidth]{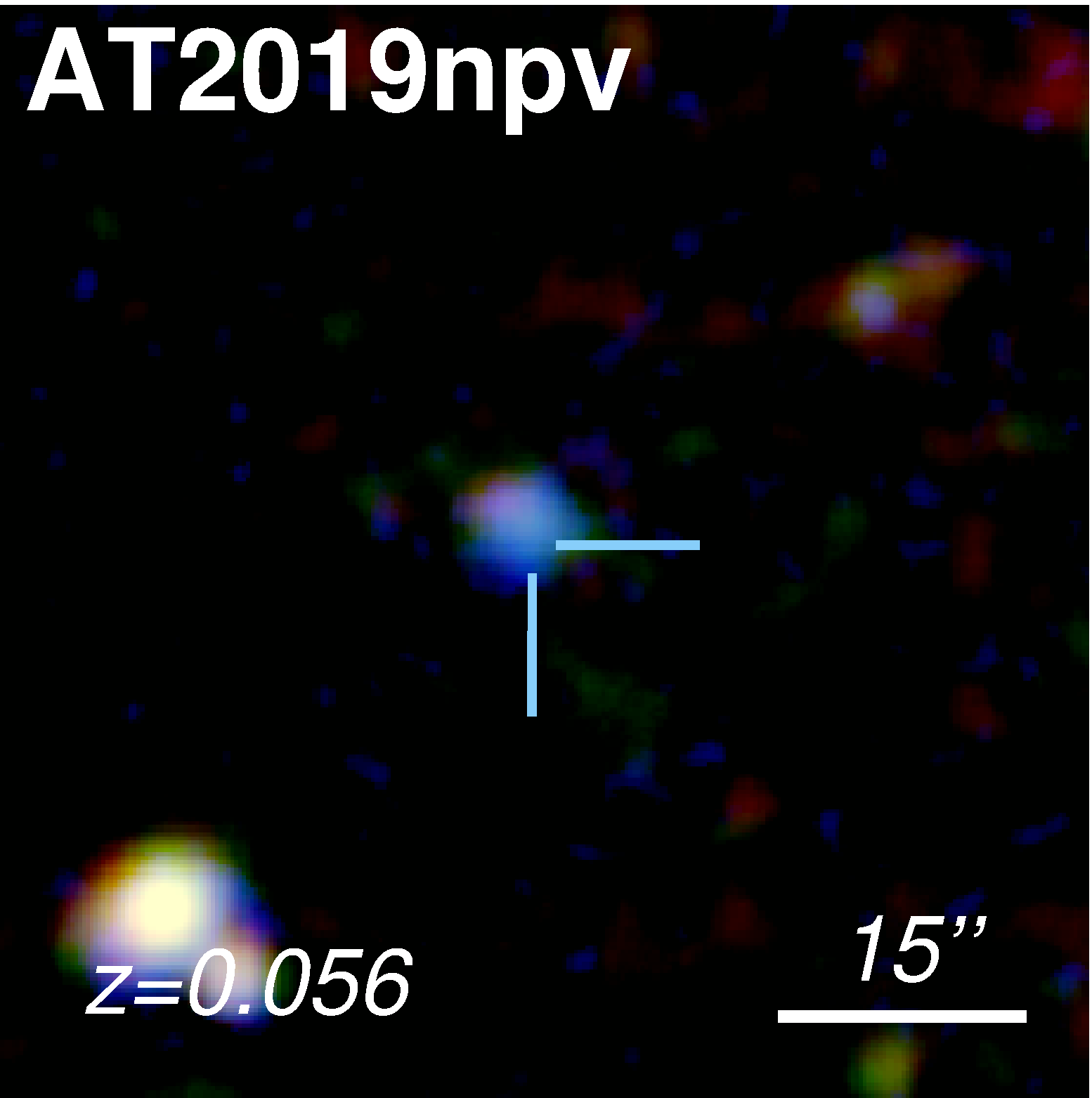}
\includegraphics[width=0.2\textwidth]{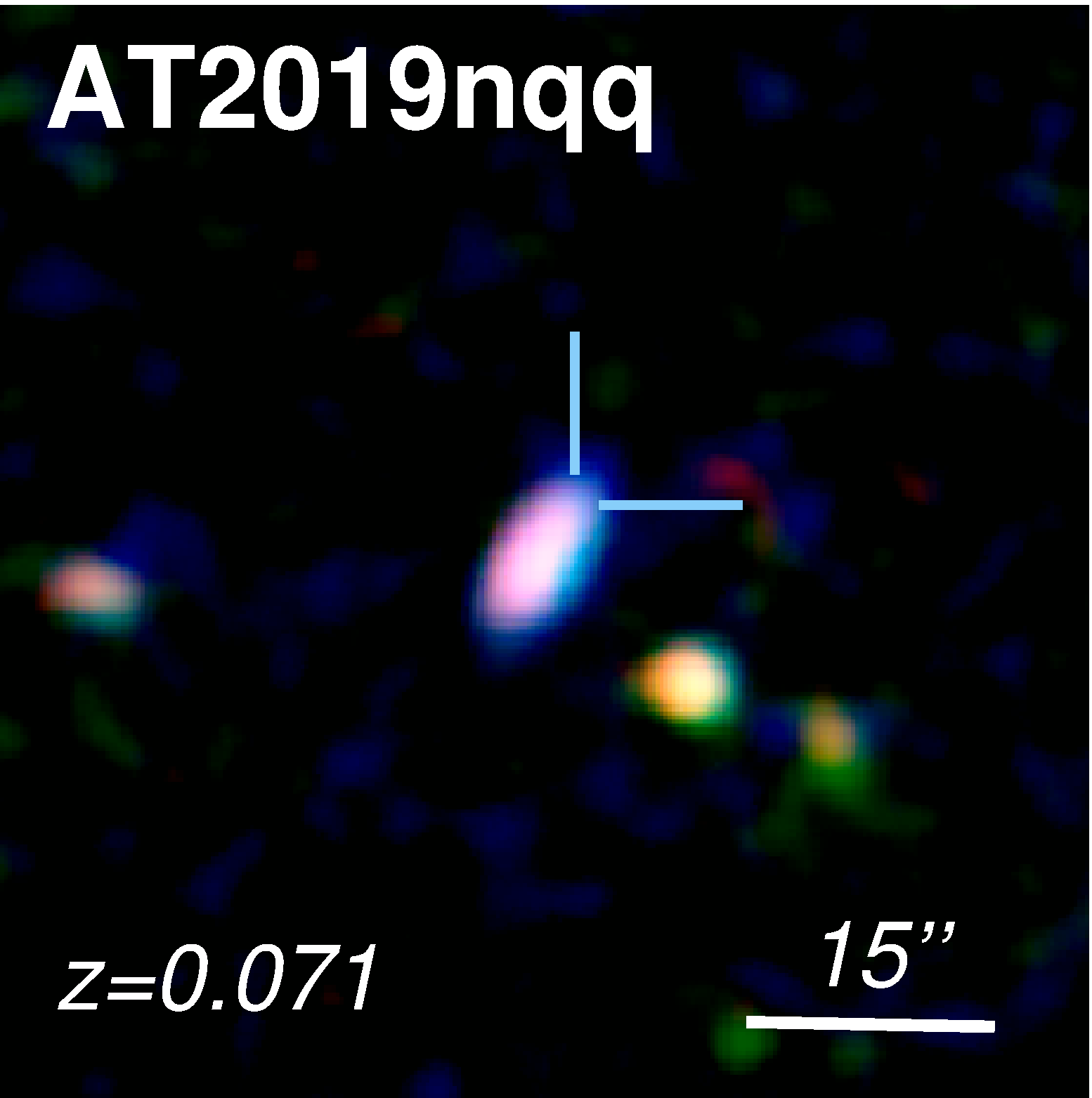}
\includegraphics[width=0.2\textwidth]{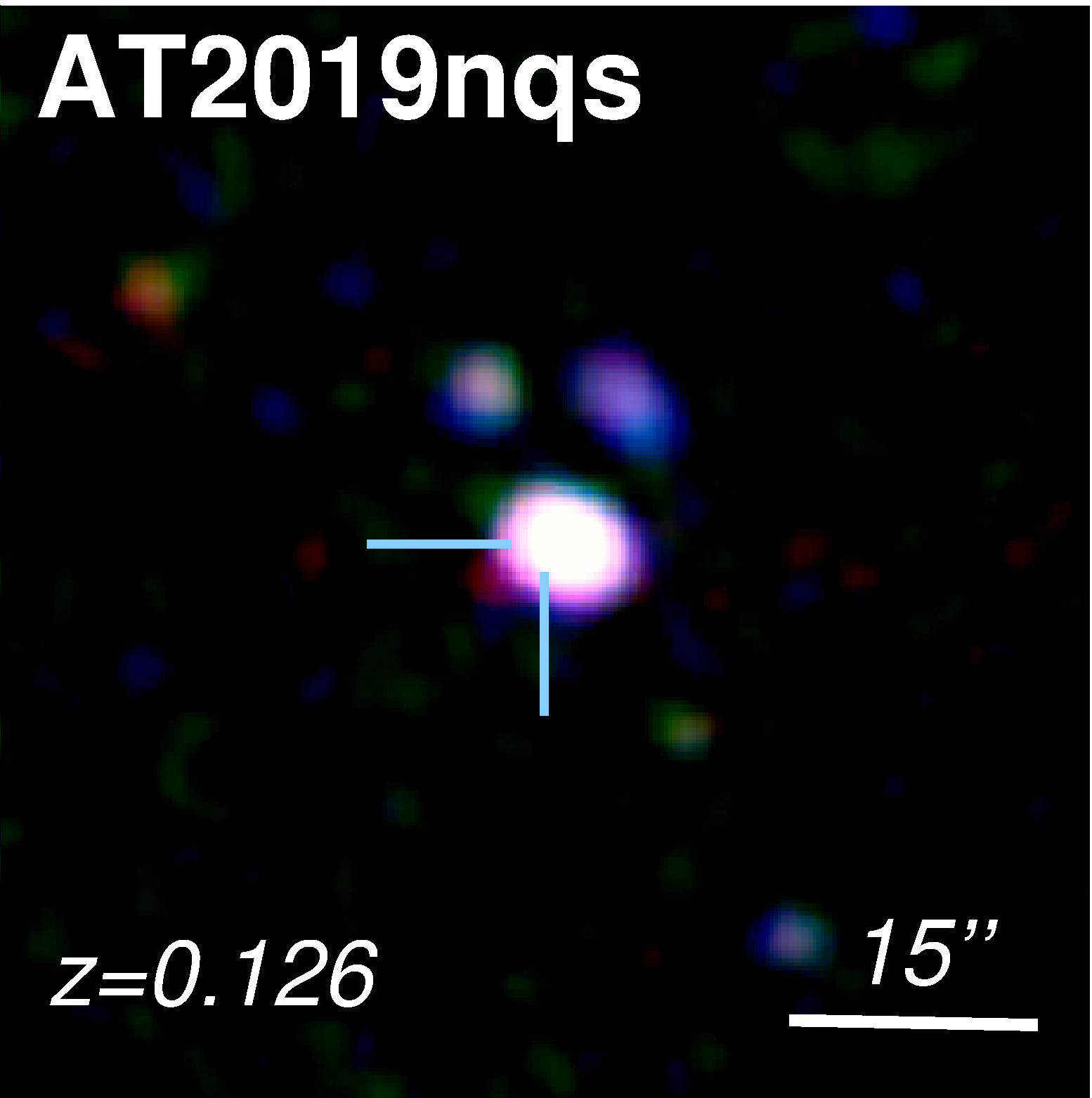}
\includegraphics[width=0.2\textwidth]{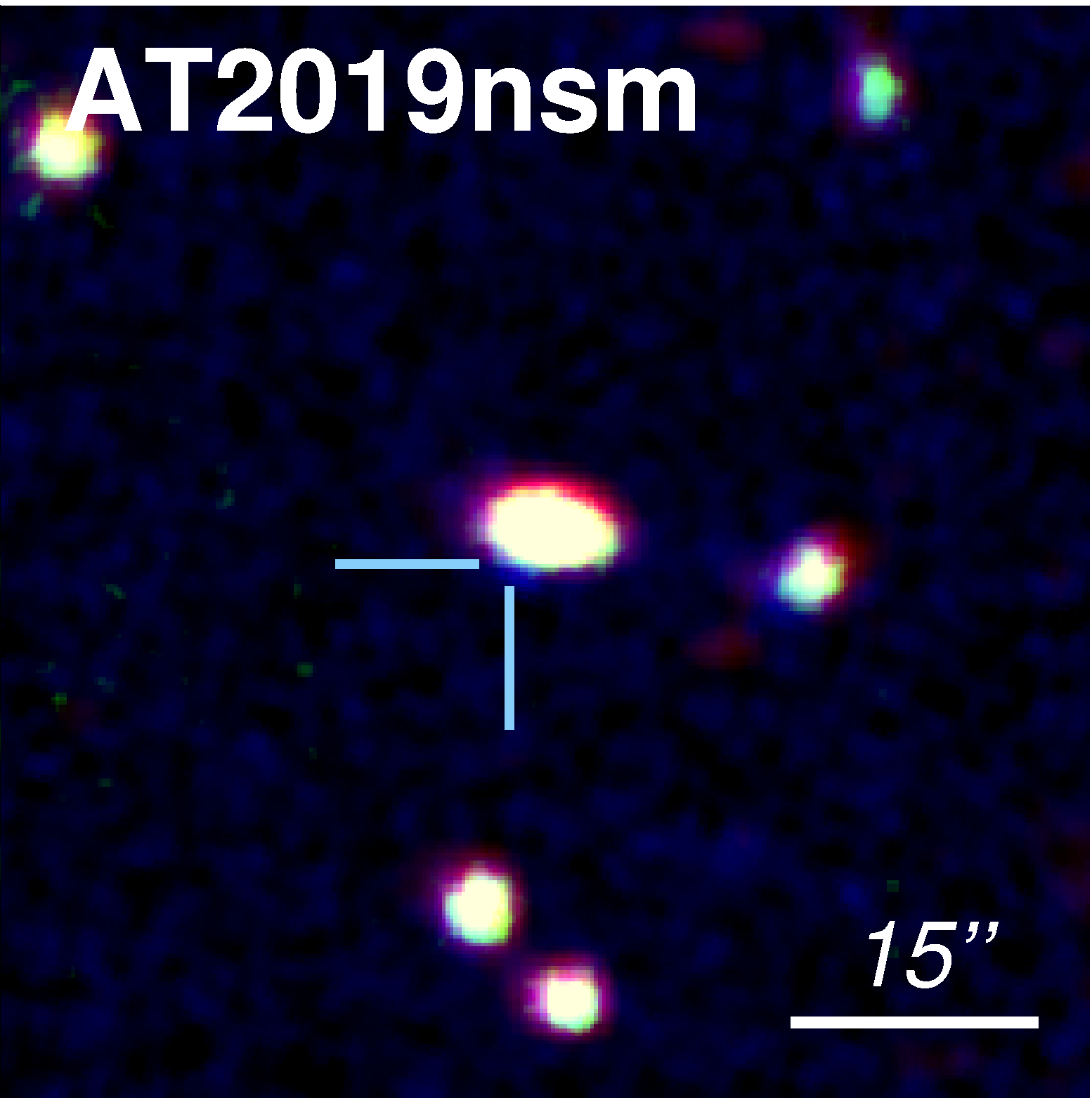}
\includegraphics[width=0.2\textwidth]{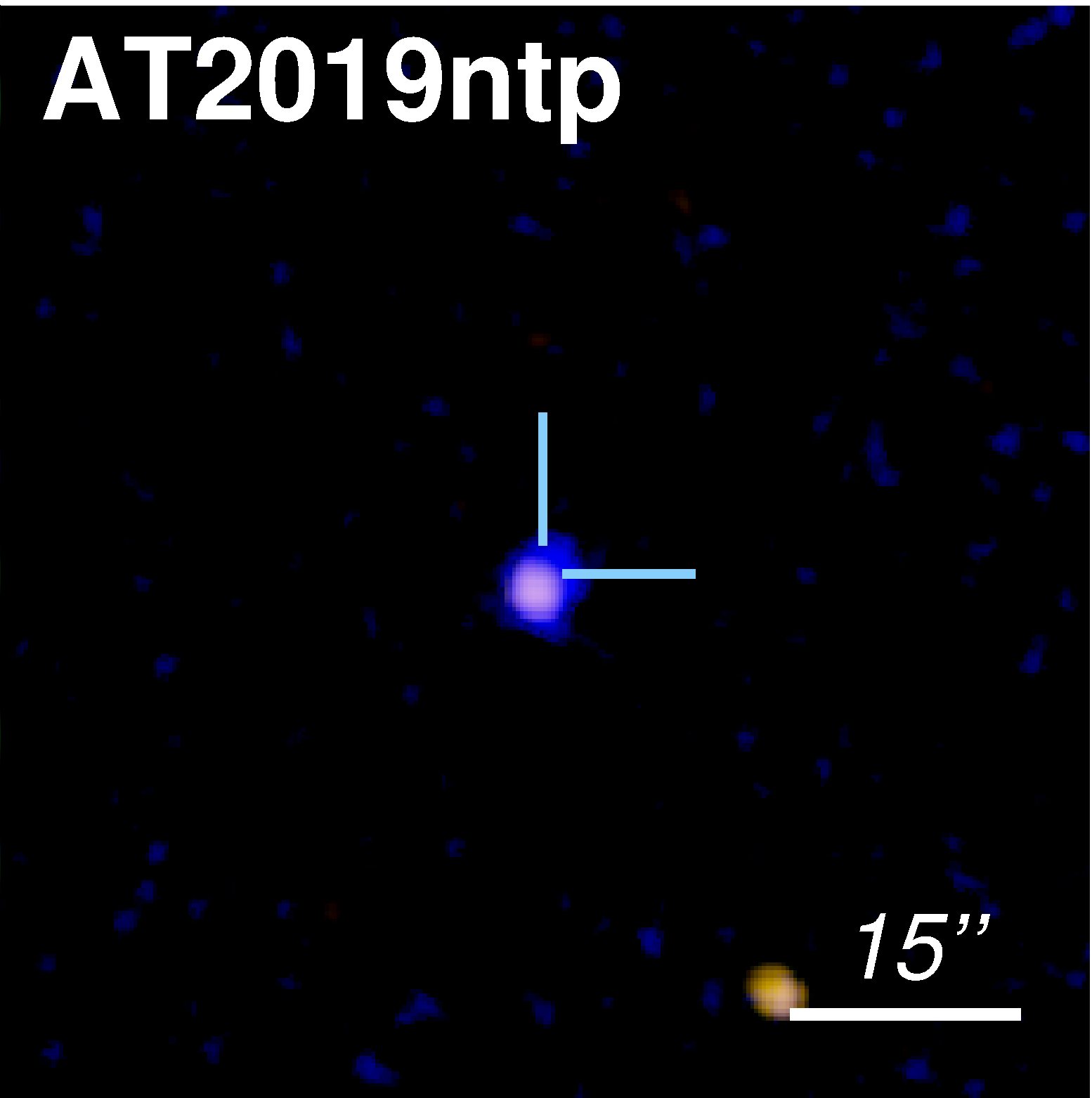}
\includegraphics[width=0.2\textwidth]{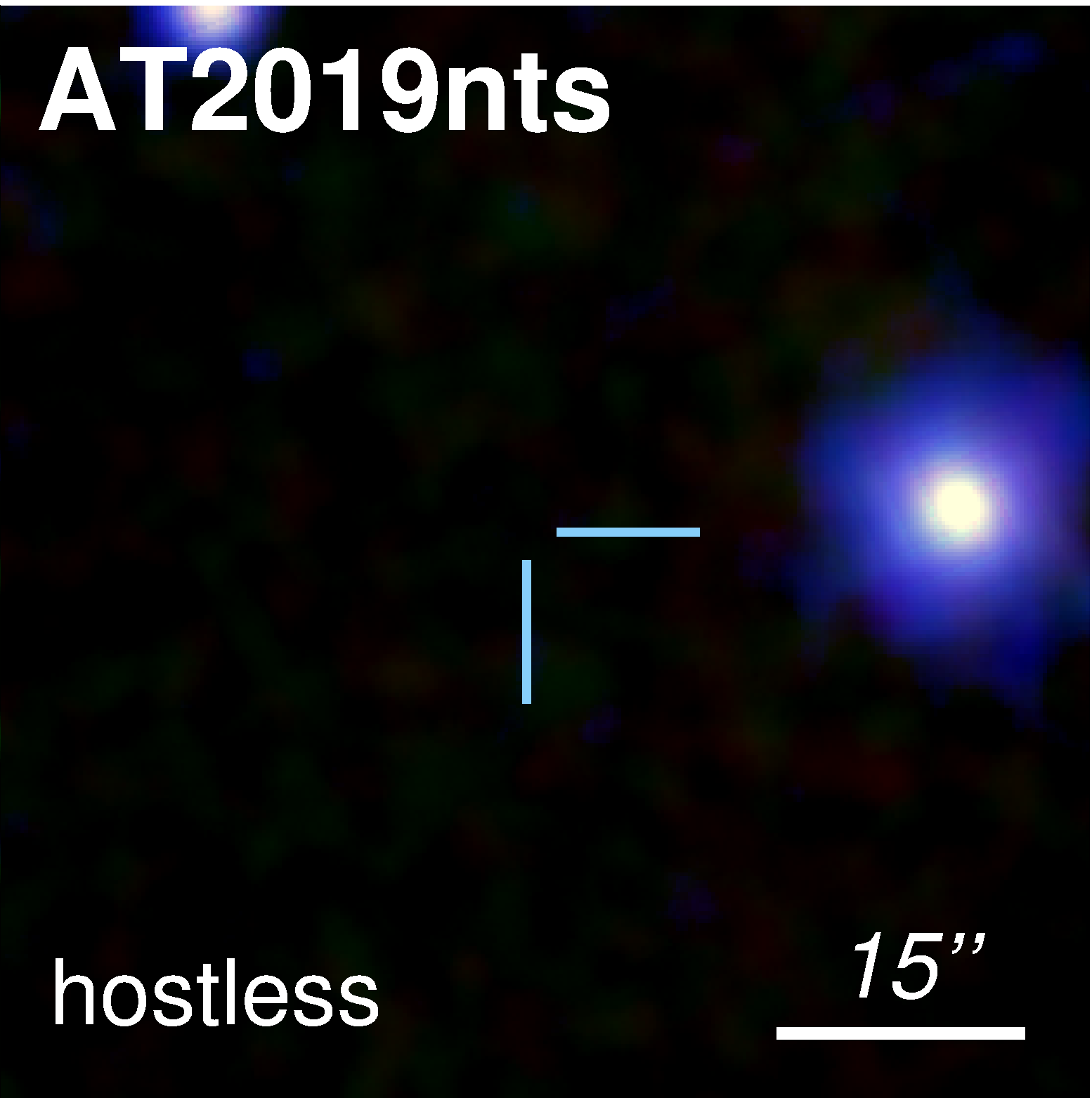}
\includegraphics[width=0.2\textwidth]{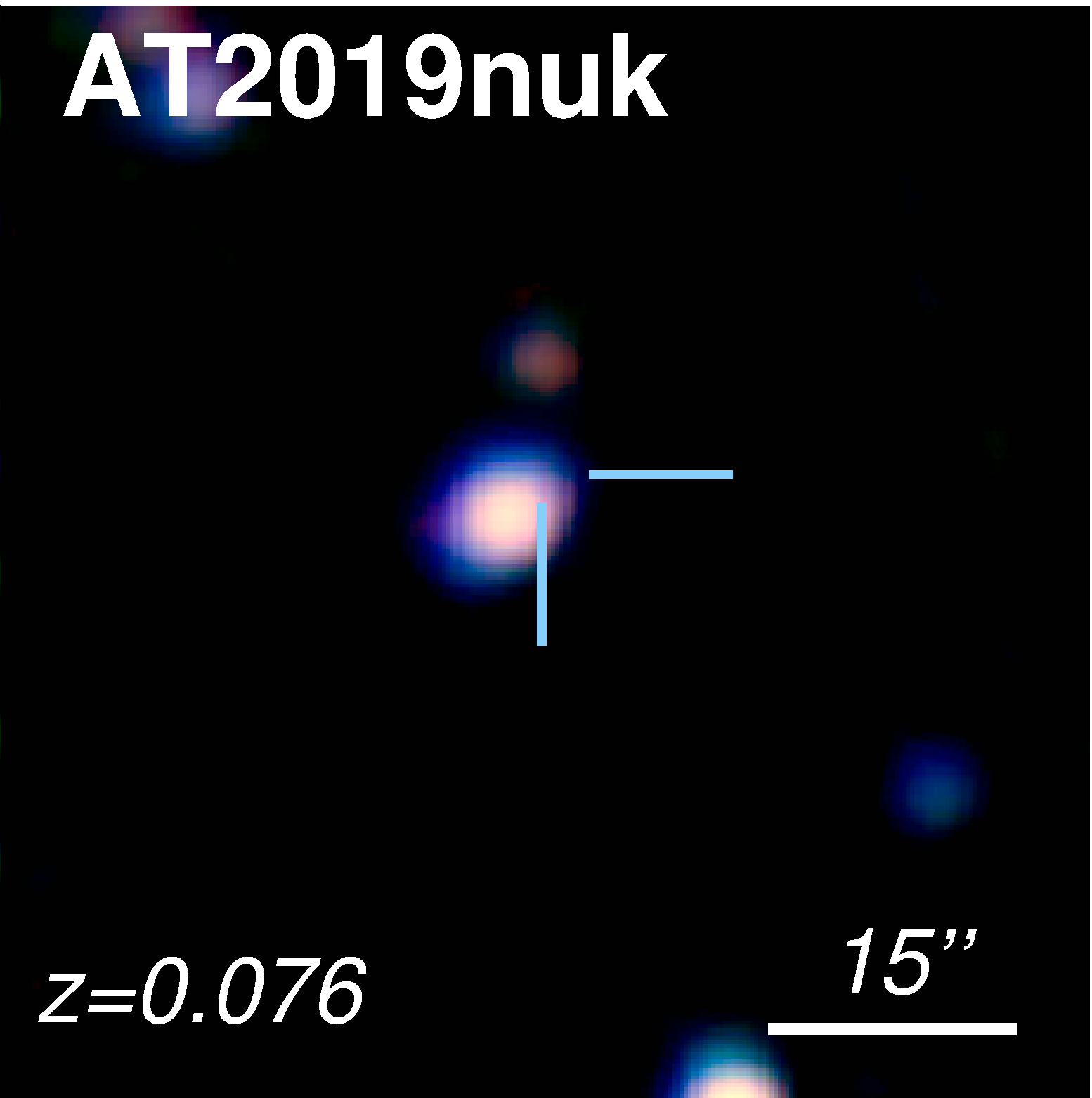}
\includegraphics[width=0.2\textwidth]{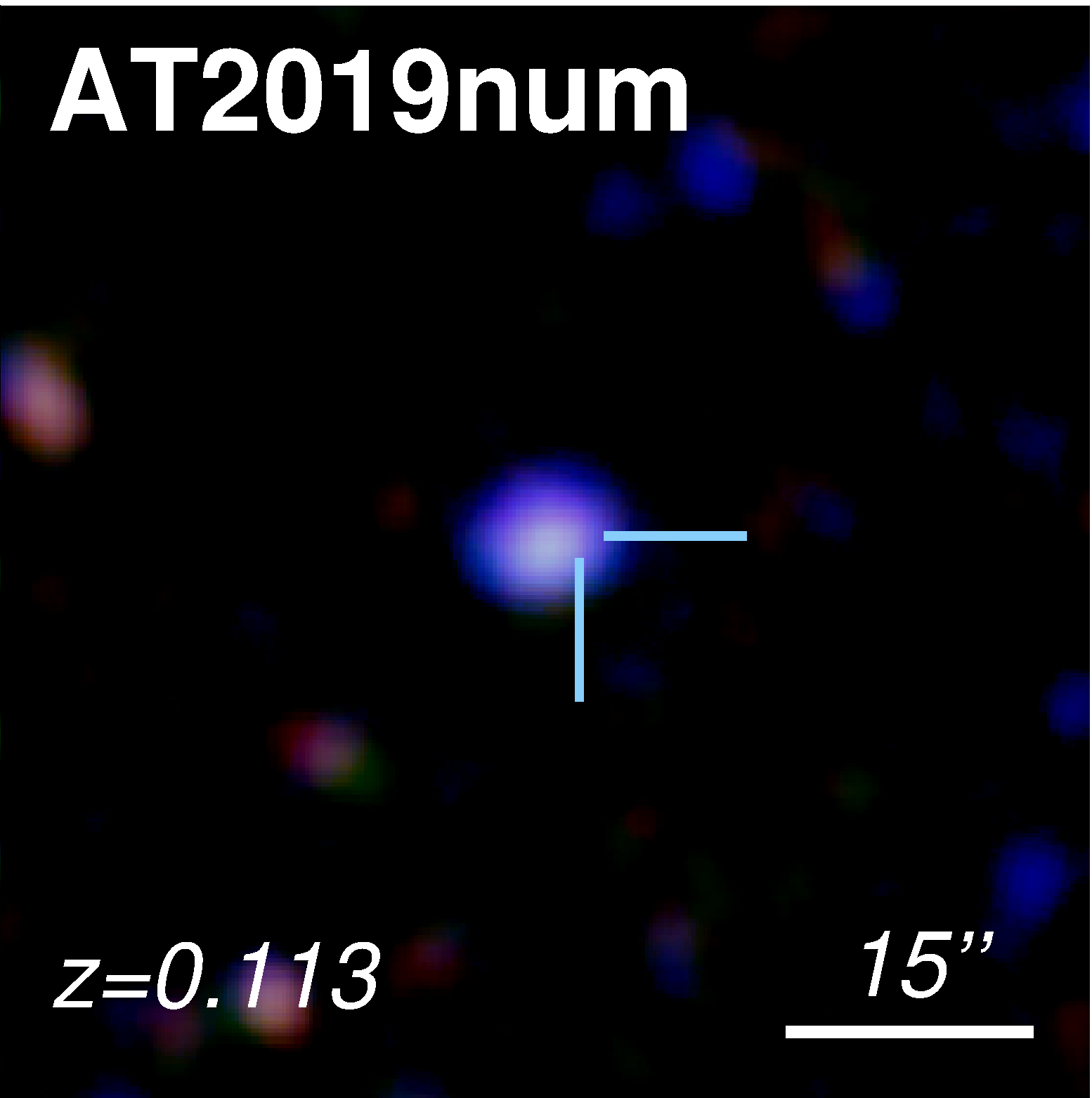}
\caption{Candidate counterparts of GW190814 observed with the RATIR camera in different filters ($r$: blue, $Z$: green, $J$: red). All images are 1.1\arcmin $\times$1.1\arcmin, and
are oriented with North up and East to the left.\label{fig:ratir}. We did not detect any of the candidates in these observations, upper limits are listed in Table~\ref{tab:ratirobservations}.
\vspace{-0.3cm}}
\end{center}
\end{figure*}

\subsection{LDT Galaxy-Targeted Search}\label{LDTGT}

We used \texttt{ligo.skymap}\footnote{https://lscsoft.docs.ligo.org/ligo.skymap/}
to cross-match the LALInference map distributed by the LIGO and Virgo Collaboration (LVC) \citep{2019GCN.25324....1L} to the Galaxy List for the Advanced Detector Era (GLADE)  v2.3 catalog \citep{GLADE}. A total of 806 (98) galaxies are identified inside the 90\% (50\%) probability volume. We targeted 14 of these galaxies, listed in Table~\ref{tab:dctobservations}, ranked according to the GW skymap probability, the B-band magnitude and the observability.    

Images were obtained using the  Large Monolithic Imager (LMI) mounted on the 4.3m Lowell Discovery Telescope (LDT) on two different nights: August, 16th and August, 18th, 2019 (1.54 and 3.54 days after the merger). 
The average airmass during the first night of observations was 2 and the seeing ranged between 1.78 - 1.95. 
On the second night the airmass varied between 2 and 2.2 and the seeing improved to values of 1.1 - 1.3.  We observed each galaxy field taking 3 exposures of 90 seconds in the $i$-band, reaching a total exposure of 270 second and an upper limit of $i$\,>\,22.9 AB mag in the field. 
The frames collected at different epochs were used to perform image subtraction and test the possible presence of variable sources (see Figure~\ref{DCTgalaxy}). Since the two images were acquired at similar epochs, our analysis is not sensitive to slowly evolving transients. Therefore, we also performed image subtraction using the Panoramic Survey Telescope and Rapid Response System (Pan-STARRS) 3$\pi$ survey images \citep[PS1;][]{PS1_Chambers2016} as a template. No transient is detected in any of the targeted galaxies. 
Derived upper limits are reported in Table~\ref{tab:dctobservations}.

For each galaxy we estimated the probability of hosting the NSBH merger
by weighting the 3D localisation probability density \citep{Singer2016} 
for the galaxy's $B$-band luminosity \citep{Gehrels2016}. 
We selected galaxies brighter than $L_B>0.1L_B^*$, where $L_B^* \approx 1.2 \times 10^{10}$ $h^{-2}$ $L_{B,\odot}$ is the characteristic galaxy luminosity of the Schechter function \citep{Schechter76}, and $h$ = $H_0$ / (100 km s$^{-1}$ Mpc$^{-1}$)$\sim$0.7 \citep{Freedman_2020}. 
Similar to \citet{2020arXiv200201950A}, our computation takes into account that the sample of GLADE galaxies inside the 90\% probability volume is $\approx$80\% complete in terms of integrated B luminosity. 
Our values therefore may differ from those reported by HOGWARTs \citep{Salmon2020}, which follows different galaxy's selection criteria and does not include the catalogue's completeness. 
Summing together the contribution of all the LDT galaxies we obtain a combined probability of $\approx$5\%, although it is difficult to make strong probabilistic statements due to the uncertainty on GLADE completeness. 

Although our observations cover a small fraction of the possible galaxies, our analysis provides an independent confirmation for the lack of candidates and it includes three galaxies not covered by other searches reported in the literature ~\citep[e.g.,][]{2020arXiv200201950A, Vieira_2020,Gomez_2019}: HyperLEDA 776957, HyperLEDA 3235869 and HyperLEDA 3235948.

\begin{table*}
 	\centering
 	\caption{LDT galaxy-targeted observations}
 	\label{tab:dctobservations}
 	\begin{tabular}{lccccccccc}
    \hline
    Galaxy Name & R.A. & Dec. & Dist.  & $M_{B}$ & $M_{K}$ & Upper Limit & Upper Limit & Probability \\
     & (J2000) & (J2000) & (Mpc) &  &  & (PS1 subtracted) &
     (LDT subtracted) & \\
     &&&&&& (AB mag) & (AB mag) & \\
    \hline
HyperLEDA-776957 & 00:53:14.256 & -25:36:49.68 & 133.2 & -19.30 & -20.53 & 21.0 & 21.5 & 0.0002\\
HyperLEDA-3235498 & 00:51:17.208 & -25:32:01.32 & 329.6 & -19.46 & -23.11 & 21.4 & 21.5 & 0.0017\\
HyperLEDA-777373 & 00:50:52.416 & -25:34:37.56  & 226.3 & -19.46 & -21.35 & 22.7 & 22.5 & 0.0035\\
HyperLEDA-3235862 & 00:53:24.864 & -25:49:36.48 & 260.3 & -19.51 & -23.71 & 21.5 & 21.4 & 0.0037\\
HyperLEDA-3235913 & 00:51:36.648 & -25:56:31.92 & 261.5 & -19.36 & -23.57 & 20.9 & 22.2 & 0.0033\\
HyperLEDA-772937 & 00:51:03.456 & -25:58:56.64 & 304.8 & -19.18 & -- & 21.5 & 21.3 & 0.0018\\
HyperLEDA-773149 & 00:51:15.768 & -25:57:39.24 &  300.6 & -19.53 & -22.53 & 21.7 & 22.0 & 0.0027\\
HyperLEDA-3235869 & 00:52:54.792 & -26:02:28.68 & 334.2 & -19.30 & -23.41 & 21.2 & 21.3 & 0.0009\\
HyperLEDA-771948 & 00:52:41.880 & -26:04:04.08 & 307.6 & -19.37 & -21.73 & 22.3 & 21.6 & 0.0019\\
HyperLEDA-3235867 & 00:52:59.016 & -26:03:03.60 & 302.0 & -19.43 & -23.05 & 22.5 & 21.3 & 0.0021\\
HyperLEDA-3235948 & 00:50:01.104 & -26:18:07.20 & 328.1 & -19.22 & -22.74 & 20.9 & 22.0 & 0.0008\\
ESO474-035 & 00:52:41.582 & -25:44:01.87 & 271.4 & -20.92 & -24.67 & 22.3 & -- & 0.0152\\
HyperLEDA-798818 & 00:50:54.447 & -23:37:54.79 & 316.8 & -21.17 & -23.79 & 21.5 & -- & 0.0056\\
HyperLEDA-2998 & 00:51:18.760 & -26:10:05.02 & 285.6 & -20.95 & -24.14 & 21.8 & -- & 0.0106\\
    \hline
    \end{tabular}
    \begin{flushleft}
    \quad 
    \footnotesize{Column 1: Galaxy name as indicated in the HyperLEDA catalog; Columns 2 and 3: Galaxy coordinates;  Column 4: distance as reported in the GLADE catalogue; Column 5: absolute 
    magnitude in the $B$-band; 
    Column 6: absolute 
    magnitude in the $K$-band; Column 7:
    95\% $r$-band upper limits derived from the subtraction of PS1 template images. 
    Column 8: 95\% $r$-band upper limits derived from the subtraction of LDT template images collected on the second epoch of observation (August 8, 18). 
    Column 9: luminosity-weighted localisation probability (see Sect. 2.2)
    }\\
    \end{flushleft}
\end{table*}

\begin{table*}
 	\centering
 	\caption{RATIR observations of candidate counterparts.}
 	\label{tab:ratirobservations}
 	\begin{tabular}{lccccccc}
    \hline
    Candidate & R.A. & Dec. & Date & $m_i$ & $m_J$ & Redshift & Spectral classification \\
    & (J2000) & (J2000) &  & (AB mag) & 
    (AB mag) & & \\
    \hline
AT2019npv & 00:53:32.316 & -23:49:58.51 & 2019-08-19 & 21.0 & 19.2 & 0.056 & SN Ib \\
AT2019ntp & 00:50:12.072 & -26:11:52.56 & 2019-08-19 & 21.2 & 19.1 & -- & SN Ic \\ 
AT2019nsm & 00:43:30.160 & -22:43:29.35 & 2019-08-20 & 21.6 & 18.9 & -- & -- \\
AT2019ntr & 01:00:01.884 & -26:42:51.59 & 2019-08-19 & 21.5 & 19.8 & 0.2 & SN II \\
AT2019nts & 00:48:31.441 & -23:06:40.80 & 2019-08-19 & 21.0 & 19.4 & -- & --  \\
AT2019ntn & 01:34:53.349 & -31:22:49.75 & 2019-08-20 & 21.3 & 18.5 & 0.1 & SN$^{a}$ \\
AT2019nuj & 00:49:01.738 & -23:14:04.93 & 2019-08-19 & 21.1 & 19.7 & 0.074$^{c}$ & --  \\
AT2019nuk & 00:54:57.827 & -26:08:04.61 & 2019-08-21 & 20.8 & 18.4 & 0.076 & -- \\
AT2019nul & 00:55:16.443 & -26:56:34.57 & 2019-08-20 & 20.8 & 18.7 & 0.098 & -- \\
AT2019num & 00:55:31.603 & -22:58:08.48 & 2019-08-20 & 20.9 & 19.2 & 0.113 & SN II \\
AT2019nun & 00:56:48.599 & -24:54:30.48 & 2019-08-21 & 21.3 & 18.7 & 0.131 & -- \\
AT2019nus & 00:44:34.557 & -22:01:44.62 & 2019-08-21 & 21.7 & 19.9 & -- & -- \\
AT2019nqc & 01:29:03.669 & -32:42:18.56 & 2019-08-22 & 20.6 & 19.8 & 0.078 & SN IIP \\
AT2019nqs & 01:33:35.164 & -31:46:48.48 & 2019-08-20 & 21.3 & 18.2 & 0.1263 & SN$^{a}$ \\
AT2019nqq & 01:23:49.217 & -33:02:04.99 & 2019-08-20 & 19.5 & 20.0 & 0.071 & SN Ic \\
AT2019osy & 00:55:47.400 & -27:04:32.99 & 2019-08-28 & 22.0$^{b}$ & -- & -- & AGN \\
    \hline
    \end{tabular}
    \begin{flushleft}
    \quad
\footnotesize{
Column 1: source identifier; 
Columns 2 and 3: source coordinates;  
Column 4: observing date; 
Column 5: 
    95\% $i$-band upper limit;
Column 6: 
    95\% $J$-band upper limit;    
Column 7: measured redshift;
Column 8: source classification derived 
from this work (see Sect.3.2), \citet{andreoni2019growth}, \citet{2020arXiv200201950A}
and from \citet{Dobie19} for AT2019osy. 
    }\\
    \footnotesize{
$^a$Uncertain type.
    }\\
        \footnotesize{
$^b$ 95\% $r$-band upper limit.
    }
    \\
        \footnotesize{
$^c$ Photometric redshift of the host galaxy
    }
    \\
    \end{flushleft}
\end{table*}

\begin{table*}
    \centering
    \caption{GTC observations log.}
    \label{tab:obslog}
    \begin{tabular}{lcccccccc}
         \hline
         Source & RA (J2000) & Dec (J2000) & Obs. Date & Exp. Time & Grism & Slit Width & Airmass & Seeing \\
         \hline
            AT2019npw & 00:56:05.742 & -25:45:01.58 & 2019-08-19 & 1x1200s & R300R & 1.23$\arcsec$ & 1.78 & 1.6$\arcsec$ \\
            AT2019nqq & 01:23:57.720 & -33:05:14.89 & 2019-08-19 & 1x1200s & R300R & 1.23$\arcsec$ & 2.11 & 2.5$\arcsec$ \\
            AT2019nqc & 01:29:03.479 & -32:45:53.50 & 2019-08-20 & 3x400s & R300R & 1.23$\arcsec$ & 2.08 & 1.8$\arcsec$ \\
            AT2019nqz & 00:46:47.397 & -24:16:32.26 & 2019-08-20 & 3x400s & R300R & 1.23$\arcsec$ & 1.66 & 1.3$\arcsec$ \\
         \hline
    \end{tabular}
\end{table*}

\begin{table*}
    \centering
    \caption{GTC/OSIRIS analysis results}
    \label{master}
    \begin{tabular}{lccc}
         \hline
         Candidate & Element (Ion) & Expansion velocity & Expected mean velocity \\
         \hline
         AT2019nqz (SNIIb) & Hydrogen (\ion{H}{I}) & -16000 km/s & -12000 km/s $^{a}$\\
         & Helium (\ion{He}{I}) & -12000 km/s & -8000 km/s $^{a}$ \\
         & & & \\
         AT2019nqq (SNIc) & Oxygen (\ion{O}{I}) & -4000 km/s & -9000 km/s $^{a}$ \\
         & Silicon (\ion{Si}{II}) & -4000 km/s & -9000 km/s $^{b,c}$ \\
         & & & \\
         AT2019nqc (SNIIP) & Hydrogen (\ion{H}{I}) & -6000 km/s &  -12000 km/s $^{a,c}$ \\
         & Helium (\ion{He}{I}) & -6000 km/s & -8000 km/s $^{a}$ \\
         & & & \\
         AT2019npw (SNIIb) & Hydrogen (\ion{H}{I}) & -10000 km/s & -12000 km/s $^{a,c}$ \\
         \hline
    \end{tabular}
    \begin{flushleft}
    \quad
    \footnotesize{
    Column 1: Transient name and its classification; 
    Column 2: Line features identified; 
    Column 3: Blueshift velocity in the reference frame of the host as measured in our analysis; 
    Column 4: Mean expected velocity obtained from the literature.
    }\\
    \quad \footnotesize{$^{a}$ \citet{Liu_2016}}\\
    \quad \footnotesize{$^{b}$ \citet{Modjaz_2016}}\\
    \quad \footnotesize{$^{c}$ \citet{Gal_Yam_2017}}\\
    \end{flushleft}
\end{table*}

\subsection{RATIR Follow-up}\label{RATIRcand}

While our DDOTI and LDT observations focused on the search of candidate counterparts, we used the 6-filter imaging camera Reionization and Transients InfraRed \citep[RATIR,][]{Butler2012,Watson2012} to monitor the sources reported by other collaborations.
Sixteen candidates discovered in the Dark Energy Camera (DECam) images and reported by the DECam-GROWTH team \citep{andreoni2019growth} and DECam-DESGW team \citep{gcn25373} were observed using RATIR (Figure~\ref{fig:ratir}). 

Observations started on August 19th (about 4.4 days after the merger) obtaining simultaneous photometry of the candidates in {\it riZJ} and {\it riYH} filters. Additional observations were collected between August, 20th and 22nd in order to characterize the sources variability, and observe newly reported candidates. Moon illumination varied from 87\% to 53\% during this period with airmass ranging between 1.7 and 3.8. The average exposure in each filter is approximately 1200~s for $r$, $i$ band, approximately 500~s for $Z$, $Y$, $J$ and $H$ band. The 2-$\sigma$ field upper limit in the $i$-band varies between 20.8 and 22.0 AB mag. On August 28th, a deep (3.8~hr) $r$-band image was obtained for the radio candidate AT2019osy \citep{Dobie19} for which we reach a field limit $r$\,$\gtrsim$\,22 AB mag.

None of the candidate counterparts were detected in our observations, the resulting upper limits are listed in Table~\ref{tab:ratirobservations}.
Optical limits are derived after subtracting the host galaxy light using PS1 reference frames. No reference frames were available for the nIR observations, and we therefore estimated our sensitivity by planting artificial point-like sources at the transient position. Our limiting magnitude is then determined by the faintest object detected with \texttt{Source Extractor} \citep{Bertin96}.

\subsection{GTC (+OSIRIS) Spectroscopy}\label{OSIRISspec}

We triggered observations of four candidate counterparts whose photometric redshifts were consistent with the distance of the GW source \citep{2019GCN.25391....1G}: AT2019nqz \citep{2019GCN.25391....1G}, AT2019nqc, AT2019nqq \citep{gcn25373}, AT2019npw \citep{2019GCN.25362....1A}. 
These observations were performed using the Optical System for Imaging and low-Intermediate-Resolution Integrated Spectroscopy \citep[OSIRIS;][]{10.1117/12.395520} spectrograph mounted on the 10.4m Gran Telescopio de Canarias (GTC). The four object spectra were obtained using the R300R grism, covering a wavelength range of 5000{\AA} - 10000{\AA}. However, we find calibration issues between 9000{\AA}-10000{\AA}, so our analysis is restricted to the wavelength range 5000 $-$ 9000{\AA}. The observations obtained with GTC(+OSIRIS) are detailed in Table \ref{tab:obslog}. 

The spectra have been reduced using standard procedures under~\texttt{IRAF/PyRAF} \citep{1986SPIE..627..733T}. They were bias subtracted and corrected for flat-field. Then, we computed for each frame a wavelength solution using an iterative method, based on previous line identifications, and applied to the observation night's lamp frames. Flux calibration was performed using standard star observations taken on the same night. The calibrated images were cosmic ray subtracted using \texttt{ccdproc} \citep{matt_craig_2017_1069648} after which the spectra were extracted from the individual calibrated science images and then combined. After extraction of the spectrum for each of the objects, we smoothed the spectrum using a Gaussian kernel to better identify broad absorption features.

We estimate the redshift of the host galaxy for each object by identifying the strongest narrow emission features in the spectrum, and assuming that they come from \ion{H}{II} regions of the galaxy. After constraining the redshift, we investigated whether the observed spectra originate from a kilonova associated with GW190814.

An important step for identifying a transient as a counterpart to a GW event is to confidently reject possible alternative origins, in particular SNe, which are a major source of contamination \citep[e.g.,][]{Cowperthwaite_2015, Doctor17, andreoni2019growth, 2020arXiv200201950A}. 

We therefore looked for SN signatures by visual inspection, as well as by matching template spectra using SNID \citep{2007ApJ...666.1024B}. For further support to our findings, we also measured the blueshift of the SN features in the reference frame of the host galaxy. We list the features that we identify and the blueshift velocities that we measure for them in Table \ref{master}. We also compared our values with the results by \citet{2014arXiv1405.1437L}, \citet{Modjaz_2016} and \citet{Liu_2016} and we report the expected mean values for the blueshift velocities in Table \ref{master}. The results of this analysis are elaborated in Section \ref{sec:specres}.

\section{Results}\label{sec:res}

\subsection{Photometric classification of candidates}

In the days following the candidate signal GW190814, over 70 candidates were proposed through GRB Circular Notice\footnote{https://gcn.gsfc.nasa.gov/other/GW190814bv.gcn3} (GCN), the majority with discovery magnitude fainter than $\gtrsim$ 21 AB in the optical. A rapid spectroscopic identification for such large number of sources is not feasible, and multi-colour imaging could more easily aid in their classification \citep[e.g.,][]{Golkhou18}. For GW190814 most of the candidates were already faint at discovery and, also due to the poor observing conditions (high airmass and moon illumination), RATIR observations did not have sufficient sensitivity to follow their temporal and spectral evolution. They can however exclude the presence of a rising light curve, typical of an off-axis afterglow \citep{Granot_2002,Ryan_2020}.

\begin{figure}
    \centering
    \includegraphics[width=\columnwidth]{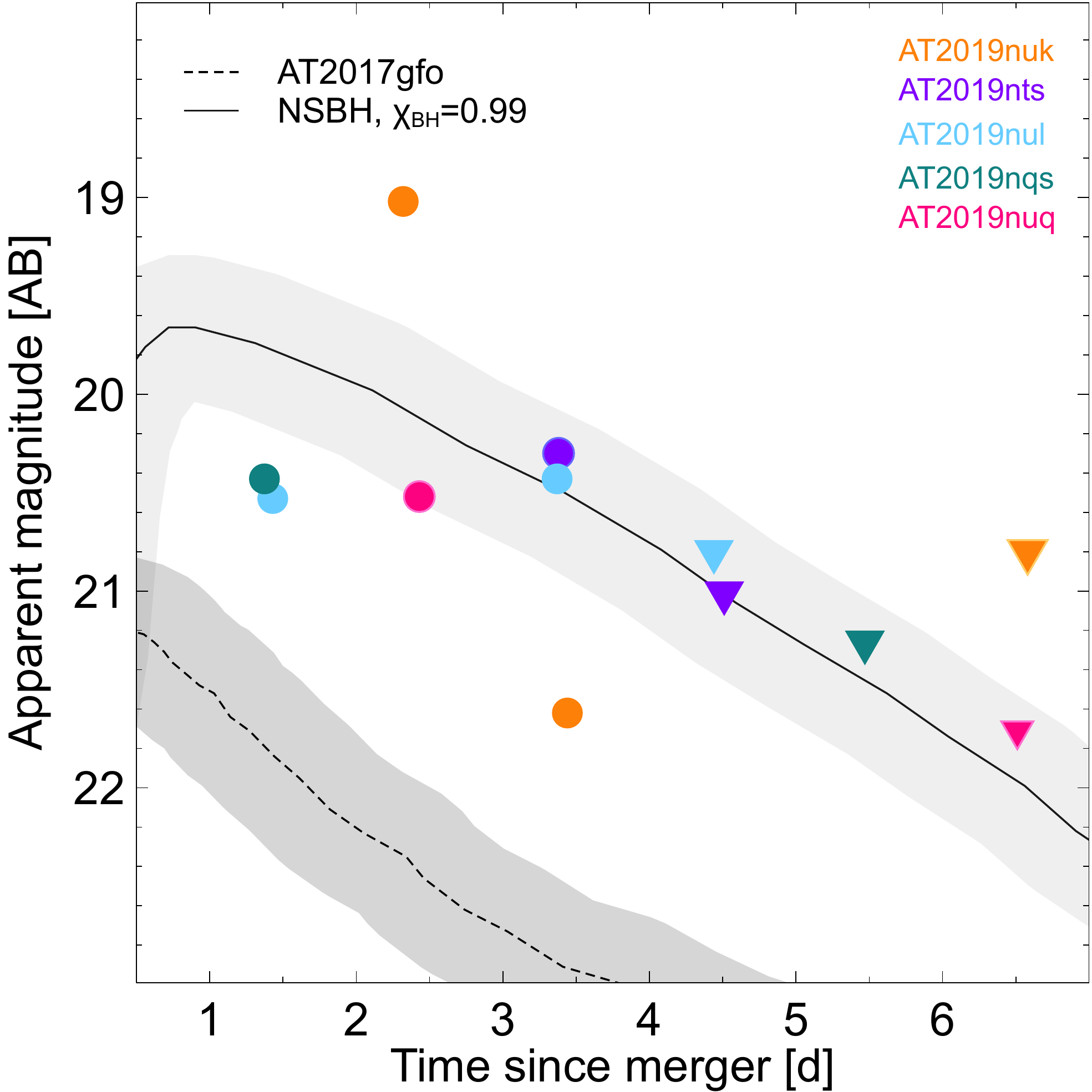}
    \caption{Temporal evolution of the brightest candidate counterparts observed with RATIR. RATIR $i$-band upper limits are marked as downward triangles. Other optical observations (marked as filled circles) were retrieved from the Transient Name Server (TNS)\protect\footnotemark. The dashed line shows the evolution of AT2017gfo (using data from \citealt{Troja2017}) shifted to 270~Mpc. The solid line shows the NSBH model from \citet{Barbieri20} for a maximally spinning BH, also shifted to 270 Mpc. The shaded areas reflect the 1\,$\sigma$ uncertainty in the source distance scale.}
    \label{fig:ratirlimits}
\end{figure}

Better constraints are possible for candidates brighter than $\lesssim$20~AB mag at discovery time. Their temporal evolution is shown in  Figure~\ref{fig:ratirlimits}.
The brightest candidate observed with RATIR is AT2019nuk.
The source, first detected with $i$\,$\sim$19.02 AB mag at 2~d, dropped to $\sim$21.6 AB mag at 3.5~d and was not detected in our observations, confirming its rapid fading. Spectroscopic observations of the host galaxy place it at $z$\,=\,0.076 \citep{2020arXiv200201950A}, consistent with the GW distance scale.
The observed temporal decay is steeper than the decay rate observed in AT2017gfo or, in general, predicted by kilonova models at a similar epoch. A sharp drop in the UV flux was observed in the early phases of AT2017gfo \citep{Evans17}, with the peak of the emission shifting toward redder wavelengths. In the case of AT2019nuk, if a rapid spectral evolution was at the origin of the optical decay, we should see its bright peak moving toward the nIR bands. As the source remains
undetected at all wavelengths (see the optical and nIR upper limits in Table~\ref{tab:ratirobservations}), we conclude that multi-colour photometry disfavours a kilonova classification. 
Furthermore, at a distance of $z$=0.076$\sim$345 Mpc, the source brightness ($M\approx$-18.7~mag at ~2 d) exceeds the expected emission from a radioactive-powered kilonova, whereas it falls within the distribution of short GRB optical afterglows. 
The rapid decay of the light curve is atypical for an afterglow too, 
although not unprecedented. For example, \citet{Piranomonte2008} observed a
similar fast decay rate $f_{\rm opt}$ $\approx$\,$t^{-5.3\pm0.9}$ for the short GRB~070707. 
Therefore, based solely on the optical/nIR follow-up, we could not unambiguously rule out the hypothesis of a (peculiar) on-axis GRB afterglow for AT2019nuk. The strongest constraint in this sense comes from the lack of a bright gamma-ray burst, ruled out by {\it Swift} observations at the time of the merger \citep{gcn25341}, 
as well as from our wide-field DDOTI observations (Sect.~\ref{jetag}).

A similar argument applies to AT2019nts, observed to fade from $i$\,$\sim$20.3 AB mag
at 4~d \citep{gcn25393} to $i$\,$>$21 AB mag at 5~d, which implies a decay slope $\gtrsim$2. 
The source lies $\approx$30\arcsec~East from a bright galaxy (Figure ~\ref{fig:ratir}), whose redshift is not known. 
By assuming the same distance of GW190814, its luminosity and decay rate would be consistent with a post jet-break GRB afterglow, while a kilonova origin appears again unlikely due to the lack of detection in the redder filters. 
The hypothesis of an on-axis GRB is however not supported by the gamma-ray and wide-field optical data. Assuming a typical afterglow decay rate to extrapolate their magnitudes back in time, both AT2019nuk and AT2019nts would have been detectable during the first night of DDOTI observations. 

Another bright candidate is AT2019nul, with a discovery magnitude of 
$i$$\sim$$20.4$ AB. 
Over the first few nights of observation, the source shows a slow temporal evolution \citep{andreoni2019growth} as well as a rather shallow spectral index, as suggested from our lack of detection in the nIR bands. These properties differ from
both kilonova and afterglow counterparts. Spectroscopic observations later published in \citet{2020arXiv200201950A} place this object at $z$=0.098, outside the 99\% probability volume of the GW source, confirming that this transient is unrelated to the merger.

\footnotetext{\url{https://wis-tns.weizmann.ac.il/search}}

The last bright candidate followed with RATIR is AT2019nqs. It was discovered on August 16th (2 d post-merger) with magnitudes of $z$=19.69 and $i$=20.43 (AB).
RATIR observations do not detect the source, which is close to its galaxy's center (Figure~\ref{fig:ratir}), and derive a limit of $i$>21.3 AB mag at 4 d.
This candidate was rapidly discarded by spectroscopic follow-up \citep{gcn25384,2020arXiv200201950A}, which placed it at a distance of  $z$ = 0.126 (about 600 Mpc), well beyond the GW distance range, and tentatively identified it as
a Type~I~SN.

For this GW event, the average sensitivity of RATIR observations
($m_i \lesssim$21 AB mag) is comparable to the peak magnitude of a AT2017gfo-like kilonova at $\approx$270~Mpc (see Fig.~\ref{fig:ratirlimits}). Therefore, they cannot exclude the presence of an event of similar brightness.
An NSBH merger involving a non-spinning BH ($\chi_{\rm BH}$=0) or a NS with a soft equation of state (EoS) would also produce a faint signal \citep{2020EPJA...56....8B}, and could not be constrained. 
Our observations are instead sensitive to the brightest kilonova 
predictions from \citet{2020EPJA...56....8B}, calculated for an NSBH merger with chirp mass $\approx$1.4 M$_{\odot}$, a stiff NS equation of state (EoS; \citealt{dd2}) and maximal BH spin $\chi_{\rm BH}$=0.99. These values differ from those derived by the analysis of the GW signal, nevertheless our comparison shows that for events at $\gtrsim$\,200 Mpc a range of merger properties could still be probed by the EM observations. Preliminary indications on the mass ratio and orientation of the merging binary would be a critical input in order to effectively target the most promising systems.

\subsection{GTC(+OSIRIS) spectral analysis results}\label{sec:specres}

\subsubsection{AT2019nqz}

\begin{figure*}
\centering
\includegraphics[scale=0.18]{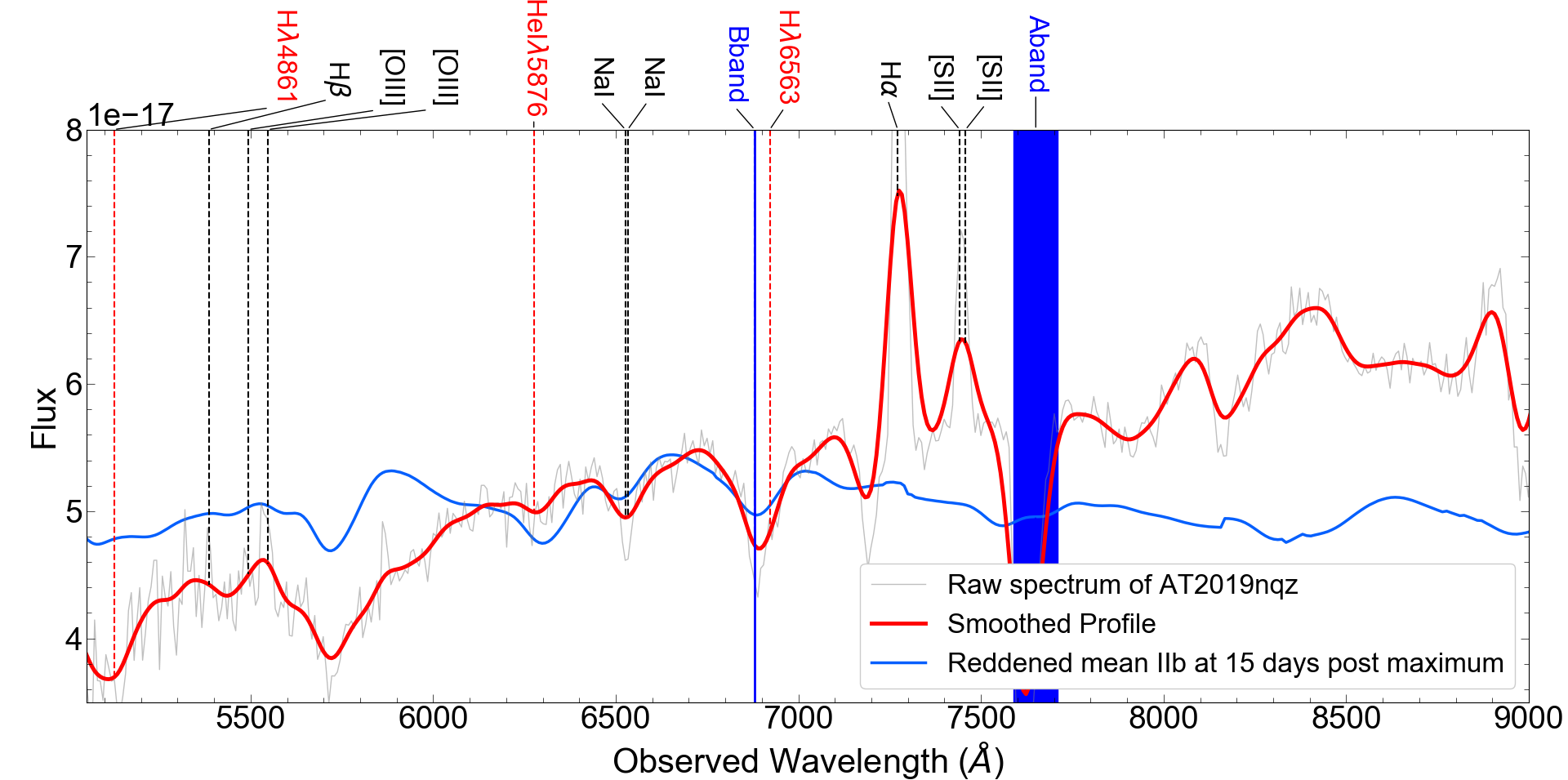}
\includegraphics[scale=0.18]{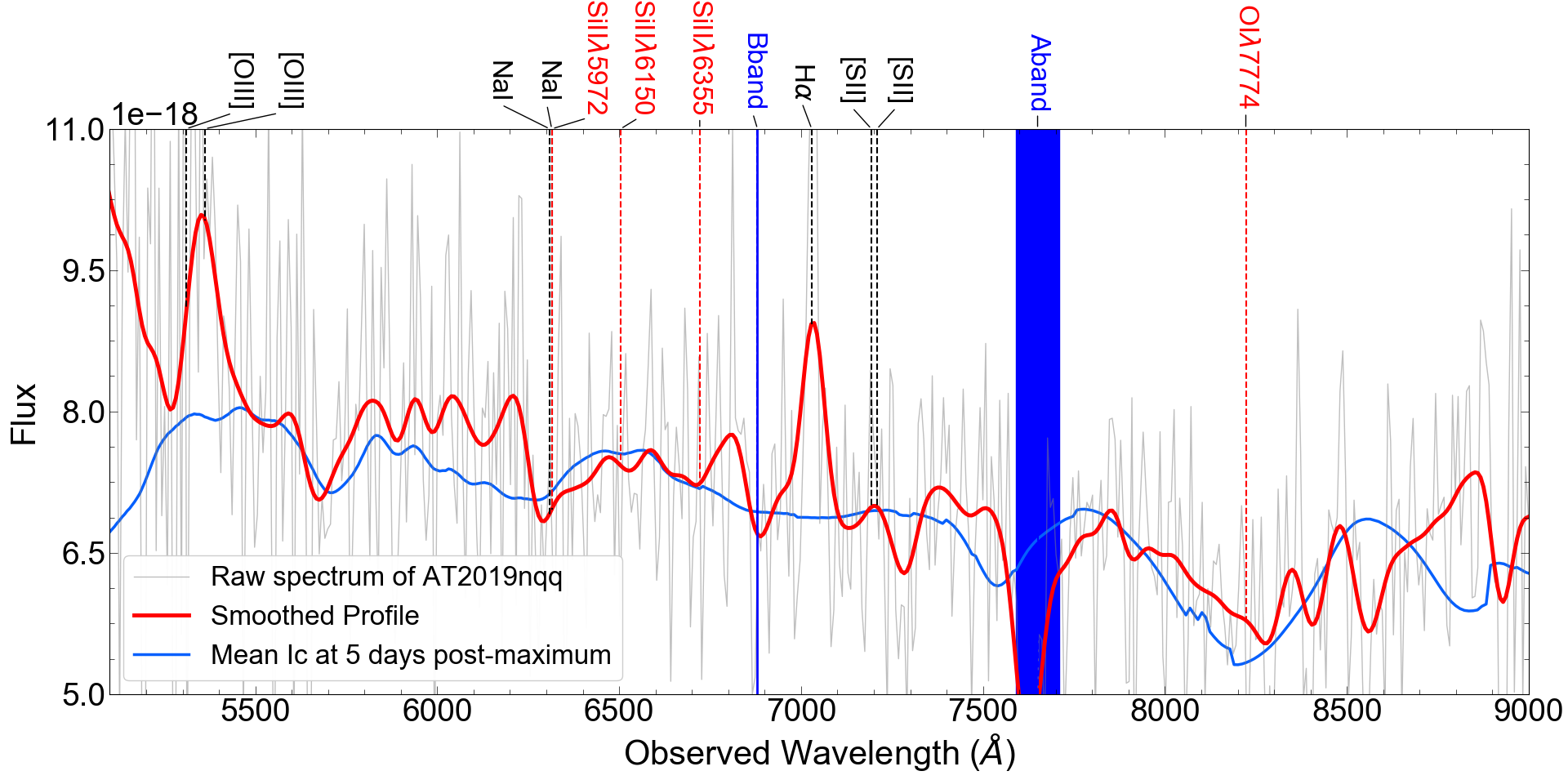}
\includegraphics[scale=0.18]{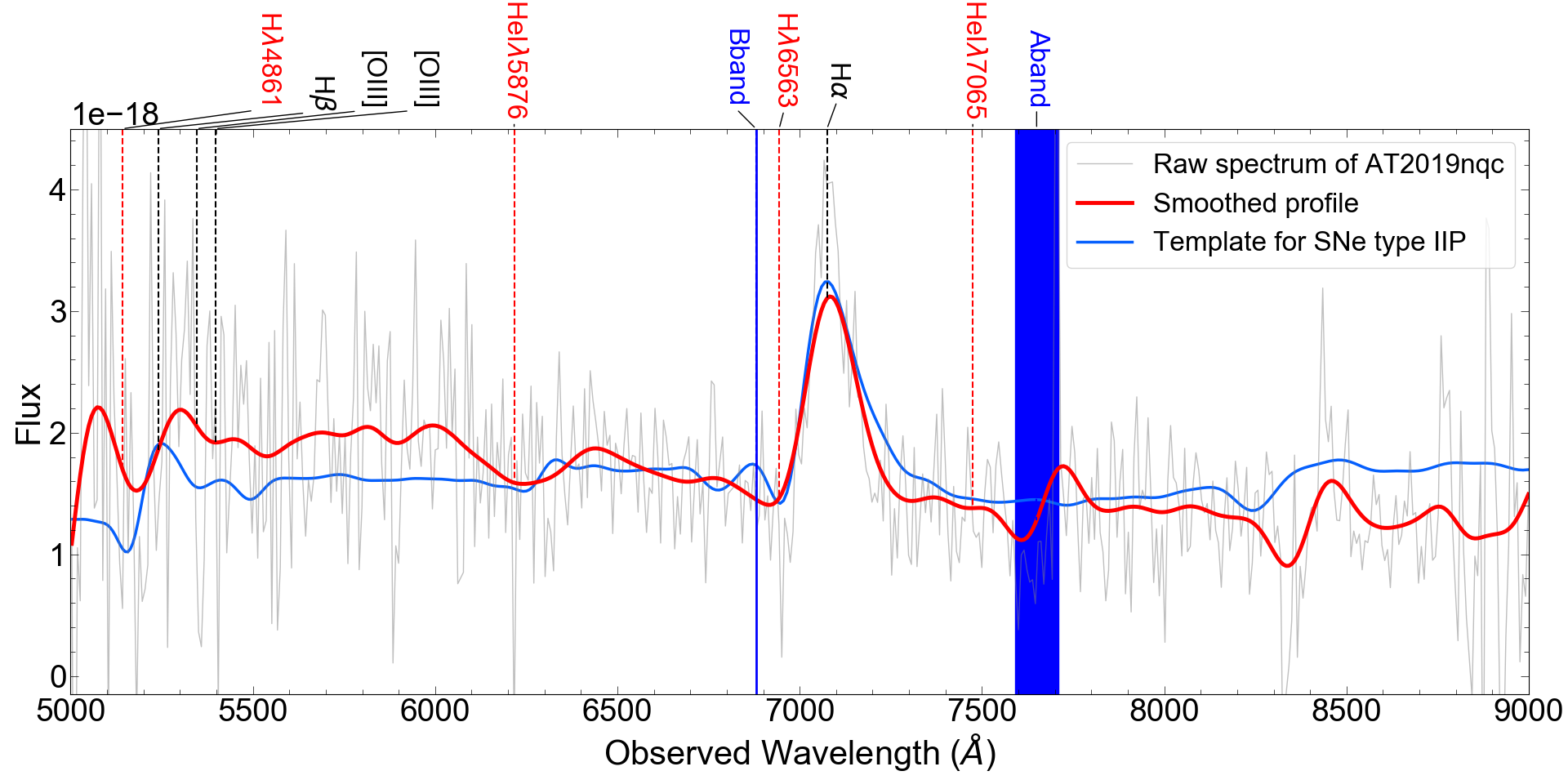}
\includegraphics[scale=0.18]{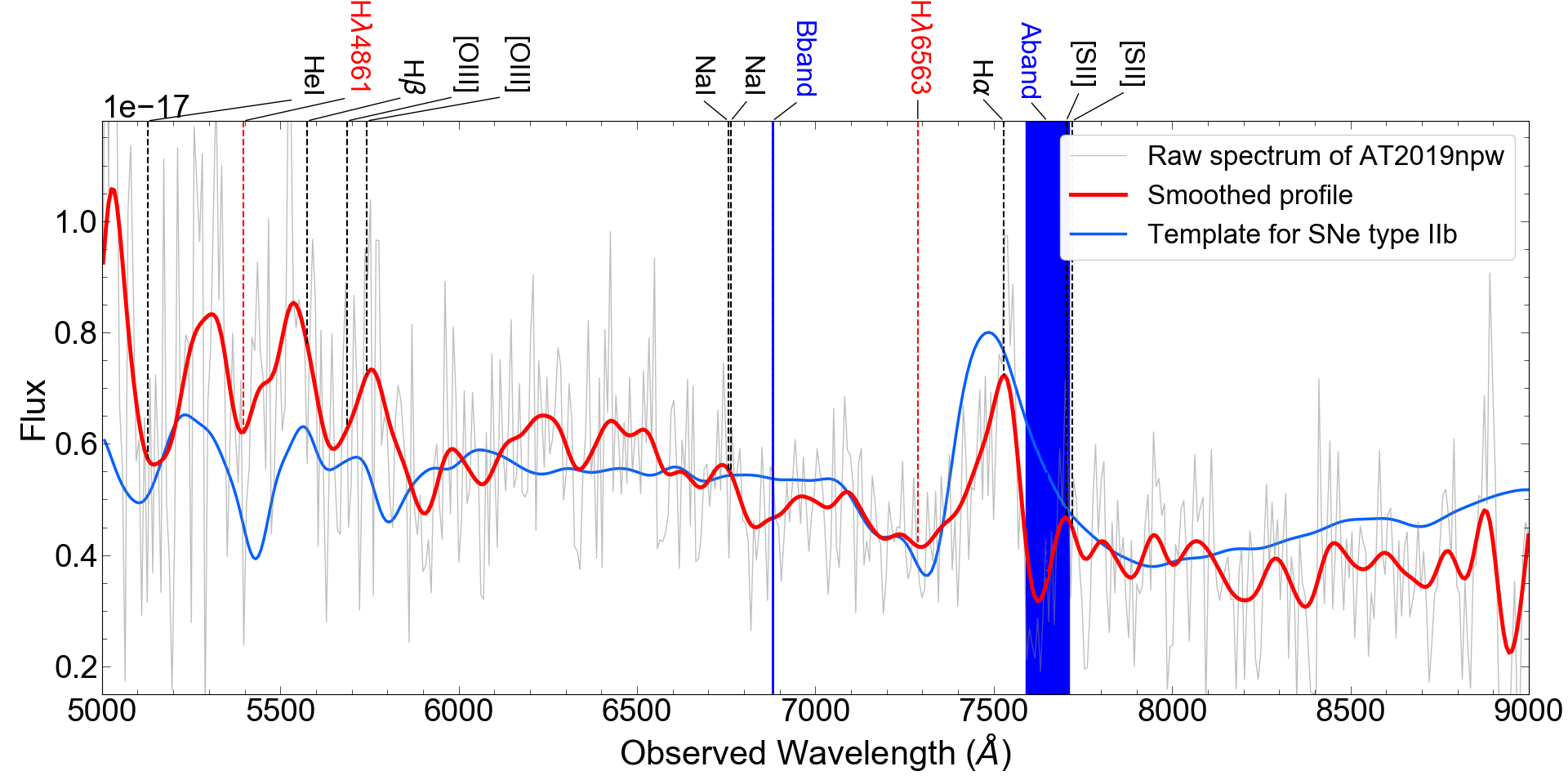}
\caption{OSIRIS spectra of AT2019nqz, AT2019nqq, AT2019nqc and AT2019npw. The GTC/OSIRIS spectrum is plotted in gray. We overplot the smoothed profile of the spectrum (red) to emphasize broad absorption features. For comparison, we also plot SNe templates obtained from the literature at the redshift of the host (blue). We mark host galaxy line features in black and the telluric bands in blue. We mark broad transient line features for which we observe a good match with the template in red.}
\label{fig:GTC}
\end{figure*}

The AT2019nqz spectrum (Figure \ref{fig:GTC}, top left panel) shows a red continuum. It has a prominent and sharp \ion{H}{$\alpha$} emission at $\sim$7273{\AA} and a sharp [\ion{S}{II}] feature at $\sim$7448{\AA}. This identification is also supported by the presence of \ion{H}{$\beta$} at $\sim$5389{\AA} and [\ion{O}{III}] emission lines at $\sim$5550{\AA}. We determine a redshift value of $z$ = 0.108 for the host spectrum using the emission features detailed above, consistent with the preliminary analysis reported in \citet[][$z$=0.1076]{ligo2019ligoc}.
This is outside the redshift range allowed by the LVC localisation, and therefore unrelated to the GW source. Nonetheless, we also attempt to classify the transient in order to better characterize the contaminants of the GW follow-up. \citet{ligo2019ligoc} also report that the transient appears to be closer than 0.5$\arcsec$ from the host. Using our low quality acquisition images obtained by GTC, we confirm that there is no evident point source distinguishable on or near the host galaxy.

The [\ion{O}{III}] and \ion{H}{$\beta$} features in the spectrum are clearly weaker than [\ion{S}{II}] and \ion{H}{$\alpha$}. This, in conjunction with the red continuum, suggests that the line of sight is strongly obscured. We confirm this by following \citet{1994ApJ...429..582C,2000ApJ...533..682C,1989agna.book.....O} and computing the Balmer decrement from the measurement of the emission line fluxes. Assuming that
\begin{equation}
    E(B-V) = 1.97 \log \frac{(\ion{H}{$\alpha$} / \ion{H}{$\beta$})_{obs}}{2.86},
\end{equation}
we obtain $E(B-V) \sim 1$. 

In order to classify the type of galaxy, and given that both AGNs and star forming galaxies are characterized by strong and narrow emission lines, we use the Baldwin-Phillips-Terlevich \citep[BPT;][]{1981PASP...93....5B} method to discern the nature of this galaxy. As we cannot measure [NII] and [OI] fluxes from our spectrum, we can only apply the BPT-SII diagnostic \citep[see][and references therein]{2006MNRAS.372..961K}. We find that
\begin{equation}
    \log \frac{[\ion{O}{III}]}{\ion{H}{$\beta$}} \approx \frac{0.72}{\log \frac{[\ion{S}{II}]}{\ion{H}{$\alpha$}} - 0.32} + 1.30,
\end{equation}
which places this galaxy on the limit between AGNs and star forming galaxies and thus cannot break the degeneracy between the two possible classifications. Therefore, as we cannot identify any point source on or near the galaxy, we cannot completely reject the possibility that this transient can be due to nuclear activity. 

However, we find that the broad absorption lines observed in the spectrum can be better explained as the superposition of SN features. We can explain the broad peaked blue absorption feature at $\sim$6926{\AA} next to the host's \ion{H}{$\alpha$} emission as a blending of \ion{H}{$\alpha$} absorption from the SN and the atmospheric B-band; with possible contribution from the host as well. We can then associate the bluest absorption at $\sim$5123{\AA} to \ion{H}{$\beta$} absorption from the SN. This is further supported by the good match of the observed absorption features with the average spectrum of a type IIb SN at 15 days post maximum obtained by \citet[][see (Figure \ref{fig:GTC}, top left panel)]{Modjaz_2016}. 
Therefore, based on these identifications and the velocities we measure (Table \ref{master}), 
we find that there is also a type II supernova in the line of sight to AT2019nqz.

The EW of the \ion{Na}{I} doublet is commonly used to infer the extinction in the line of sight to SNe \citep[e.g.,][]{1990A&A...237...79B,2003fthp.conf..200T,2018MNRAS.478.3776D} and galaxies \citep{2012MNRAS.426.1465P}, despite the fact that this method has known limitations \citep[see, e.g., ][]{2011MNRAS.415L..81P}. From our spectrum, we measure a rest frame EW(\ion{Na}{I})\,=\,3.97\AA. This large value is not seen in the SDSS galaxy sample collected by \citet{2012MNRAS.426.1465P} (we expect EW $\sim 0.6${\AA} for $E(B-V)\sim1$ from their Fig. 8). However, our values are consistent with the relationship found for SNe as inferred from Fig. 3 of \citet[][]{2003fthp.conf..200T} and Fig. 1 of \citet[][]{2011MNRAS.415L..81P}. Therefore, the EW(\ion{Na}{I}) value we measure from our spectrum strengthens our previous result that AT2019nqz is a type II supernova.

\subsubsection{AT2019nqq}

The AT2019nqq spectrum shows a blue continuum (Figure \ref{fig:GTC}, top right panel). We calculate a redshift of $z=0.071$ for the host. Our result is consistent with the value reported by \citet{andreoni2019growth}, and places this transient within the distance range of the GW source. 
The host galaxy's redshift is constrained using a prominent and narrow emission line at $\sim$7032{\AA}, which we interpret as \ion{H}{$\alpha$} emission from the host. This identification is supported by the [\ion{O}{III}] emission features at $\sim$5230{\AA}. We also marginally detect an emission feature at $\sim$7195{\AA}, which is consistent with [\ion{S}{II}] doublet emission at the same redshift. At difference with \citet{andreoni2019growth}, we do not find \ion{H}{$\alpha$} emission wide enough to support their Type II SN classification. Furthermore, the apparent P-Cygni profile is most likely due to the atmospheric B-band.

Instead, the absence of strong hydrogen features favours a type I SN classification. We identify a weak \ion{Si}{II} absorption feature centered at $\sim$6700{\AA}. We also identify an absorption feature centered at $\sim$8198{\AA} which is consistent with \ion{O}{I} absorption from the SN. The combination of the weaker \ion{Si}{II} relative to the \ion{O}{I} feature and velocity values favours a type Ic classification \citep[velocity values are summarised in Table~\ref{master}; see][]{Modjaz_2016,Gal_Yam_2017}.
We overplot in (Figure \ref{fig:GTC}, top right panel) an average type Ic spectrum 5 days post maximum for comparison with our spectrum, finding a reasonable match between them, keeping in mind that our spectrum is not host-subtracted. We thus find that the AT2019nqq spectrum favours a type Ic classification.

\subsubsection{AT2019nqc}

The AT2019nqc spectrum (Figure \ref{fig:GTC}, bottom left panel) shows a flat continuum. The host redshift is constrained assuming that the prominent peak component at $\sim$7536{\AA} is \ion{H}{$\alpha$} coming from \ion{H}{II} region(s) in the galaxy, 
at a redshift of $z$\,$\sim$0.078, consistent with the measurement of \citet{andreoni2019growth},
and within the 95\% distance range of GW190814.

We do not detect convincing [\ion{O}{III}] or \ion{H}{$\beta$} features, suggesting that the spectrum is dominated by a transient source. This is supported by the presence of a very broad \ion{H}{$\alpha$} emission component next to a weak absorption feature. These characteristics are typical of type II SNe.

The transient spectrum has a \ion{H}{$\alpha$} P-Cygni profile. We also detect \ion{H}{$\beta$} absorption at $\sim$5131{\AA}, and \ion{He}{I} absorption features at $\sim$6223{\AA} and $\sim$7463{\AA}. We note that the feature at $\sim$7463{\AA} is very weak. These characteristics suggest a type II SN classification for this spectrum, as independently suggested by \citet{andreoni2019growth}. For comparison, we overplot the spectrum of the type IIP SN2005cs \citep{muendlein2005supernova}. The template spectrum is at 4 days post maximum.

\subsubsection{AT2019npw}

The AT2019npw spectrum (Figure \ref{fig:GTC}, bottom right panel) shows a blue continuum. The host redshift is constrained using the prominent \ion{H}{$\alpha$} line at $\sim$7536~{\AA} and [\ion{O}{III}] emission features at $\sim$5750{\AA}. We further identify [\ion{S}{II}] emission at $\sim$7708{\AA}. We find the redshift of the host to be $z$=0.147, well beyond the GW distance range. 

This object has been reported by \citet{andreoni2019growth} as a type IIb SN and the combination of spectral features that we identify supports this conclusion.

The transient spectrum has a clear broad \ion{H}{$\alpha$} absorption feature at $\sim$7297{\AA}. We further find \ion{H}{$\beta$} absorption at $\sim$5423{\AA}. The velocity values we measure for this spectrum are summarised in~Table~\ref{master}

We plot the type IIb SN2004et \citep[see][]{zwitter2004supernova} as a reference spectrum for comparison. The spectrum is at 9 days post maximum. We find good agreement in the profile of our spectrum and template, which supports our classification. 

\section{Discussion}\label{sec:dis}

\subsection{Optical candidates follow-up campaign}\label{sec:Opticalfollowup}

\begin{figure}
\centering
\includegraphics[width=\columnwidth, trim=10 0 5 0, clip]{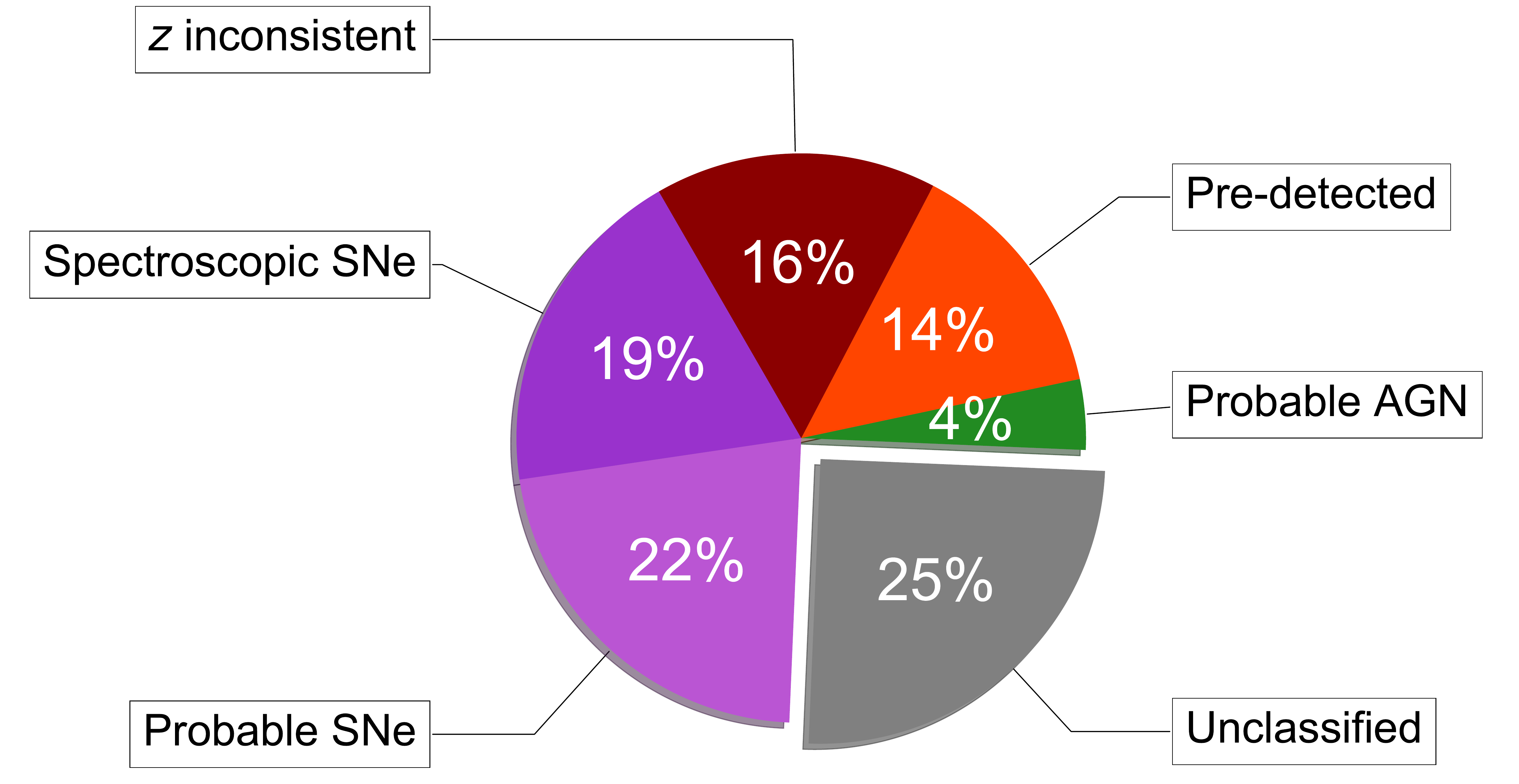}
\caption{Classification scheme for candidate optical counterparts of GW190814. Archival detections, Solar System objects and proper motion star are grouped under pre-detected. Sources ruled out on the basis of redshift, either photometric or spectroscopic, are combined under $z$-inconsistent. Probable SNe and probable AGN were classified on the basis of their photometric evolution.  }
\label{fig:Opticalpiechart}
\end{figure}

A total of 85 optical transients, with brightness ranging between 18 and 24 AB mag,  were identified as candidate counterparts of GW190814. Of these, 71 were announced via Gamma-ray Coordinates Network notices (GCNs) and 14 were reported at a later time through publications. 
Figure~\ref{fig:Opticalpiechart} summarizes the results of 
this community wide effort. 
Based on the results of our analysis and other works \citep[e.g.,][]{andreoni2019growth,2020arXiv200201950A},
42 candidates can be ruled out as counterparts of the GW source: 
17 are spectroscopically classified, 5 have photometric redshifts inconsistent with the GW signal, 9 are associated to a host galaxy with redshift inconsistent with the GW signal, 9 have archival detections, and 2 are moving objects.
This then leaves 43 candidates ($\sim50\%$)
without a secure classification.
For these, 19 are probable SNe based on their light curve evolution, and 3 are probable Active Galactic Nuclei (AGN). 
The remaining 21 sources ($\sim$25\%) are left unclassified.
Of these, 10 candidates have photometric redshifts consistent with the GW distance (within the 95\% uncertainties), 2 have host galaxy spectra whose redshifts are consistent with the GW distance, and 9 have no constraints on their distance scale. 
The magnitude of the candidates that are redshift consistent with the GW distance ranges from 23.2 to 21.3 AB mag.

We find that of the 17 spectroscopically classified candidates, one is a proper motion star and 16 are identified as SNe:
9 are Type II, 2 are Type Ibc, 2 are Type Ia and 3 have an unclear sub-type classification. Combining the number of classified SNe and probable SNe, we get a total of 35 optical candidates ($\approx$40\%) that could be classified as SNe. This is consistent with the predictions that SNe are major contaminants in the optical follow-up of GW events due to their rates and luminosity \citep[][]{Nissanke13,Cowperthwaite_2015}.

\begin{figure}
\centering
\includegraphics[width=\columnwidth]{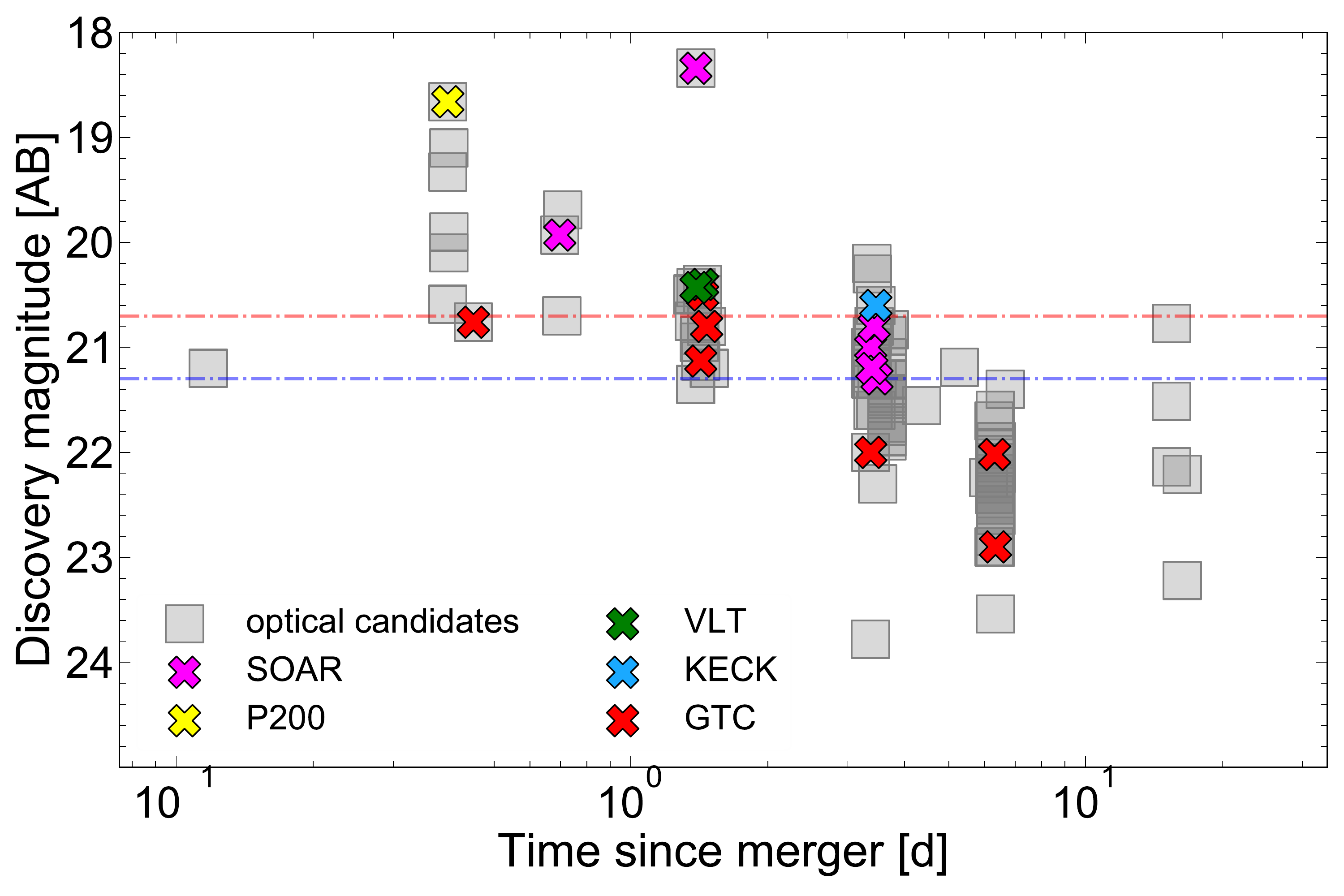}
\caption{Discovery magnitude of optical counterpart candidates for GW190814 as a function of time. The discovery magnitudes are as obtained from GCNs. Spectroscopically classified transients are coloured differently to highlight the facilities used for the classification. The median magnitude for the entire set of 85 candidates (21.3 mag, blue) and the median magnitude for the subset of spectroscopically classified candidates (20.7 mag, red) are marked as horizontal dashed lines.}
\label{fig:Opticalcandidates}
\end{figure}

We further investigate whether the results may be affected by observing biases, preferentially targeting a particular type of transient. 
Figure~\ref{fig:Opticalcandidates} reports the discovery magnitude of all the proposed candidates as a function of their time of announcement. 
On average, brighter sources were reported at early times. 
We calculate a median discovery magnitude of 21.3 for the entire sample of candidate counterparts. The median magnitude for the subset of spectroscopically classified candidates is 20.7, only slightly brighter than the complete sample.
We use the Kolmogorov-Smirnov test to compute the probability that the two sets of magnitude can be drawn from the same probability distribution finding a p-value of 0.08. Therefore, we cannot identify any significant difference between the entire set of candidates and the subset of spectroscopically classified sources.

The follow-up of candidates reported within the first four days was very thorough: 27 candidates were announced via GCNs, out of which 18 ($\sim67~\%$) have spectroscopic observations (13 with a spectroscopic classification). In a few cases (AT2019nqq, AT2019nqc and AT2019npv) multiple spectroscopic observations were reported. For the candidates announced at later times ($>$4~d), we do not recognize any clear pattern in the selection criteria for spectroscopic follow-up. Instead, we note that most of these candidates remain unclassified. Therefore, time rather than brightness was the discriminant factor in obtaining a spectroscopic identification. 

\begin{figure*}
\centering
    \includegraphics[width=\columnwidth]{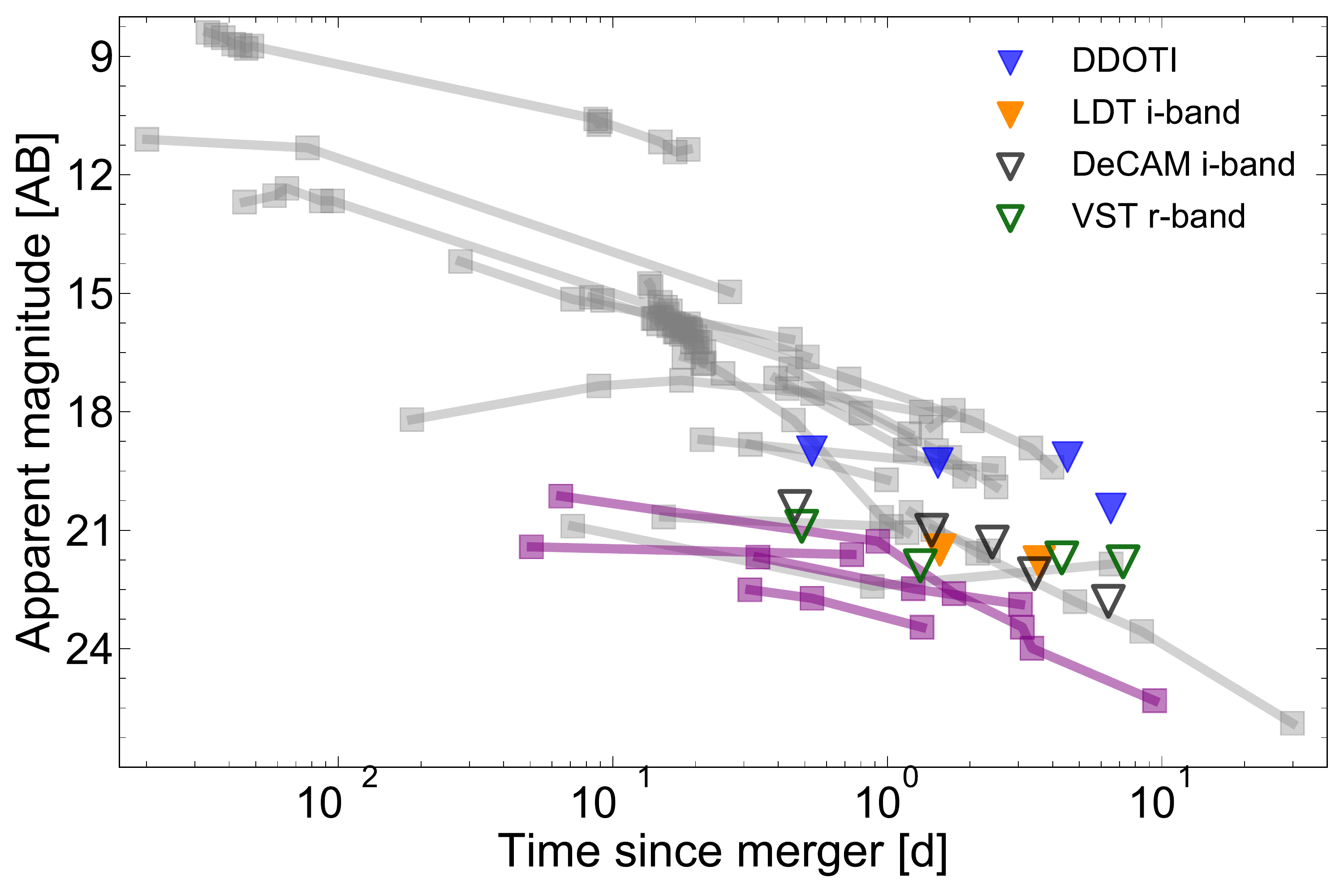}
    \hspace{0.2cm}
    \includegraphics[width=\columnwidth]{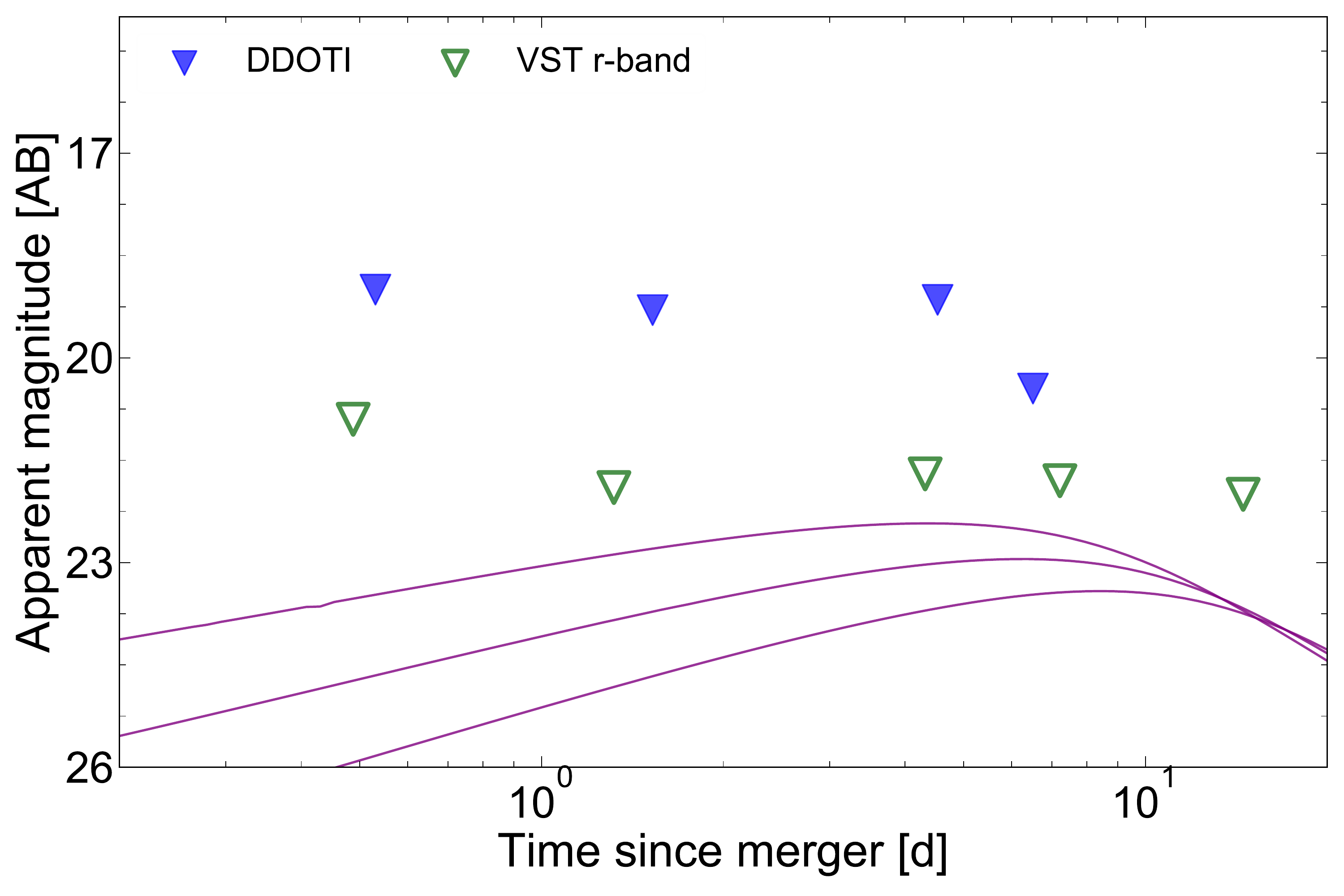}\llap{\makebox[0pt][l]{\hspace*{-3.8cm}\raisebox{1.18\height}{\includegraphics[trim=30 25 5 5 , clip, scale=0.14]{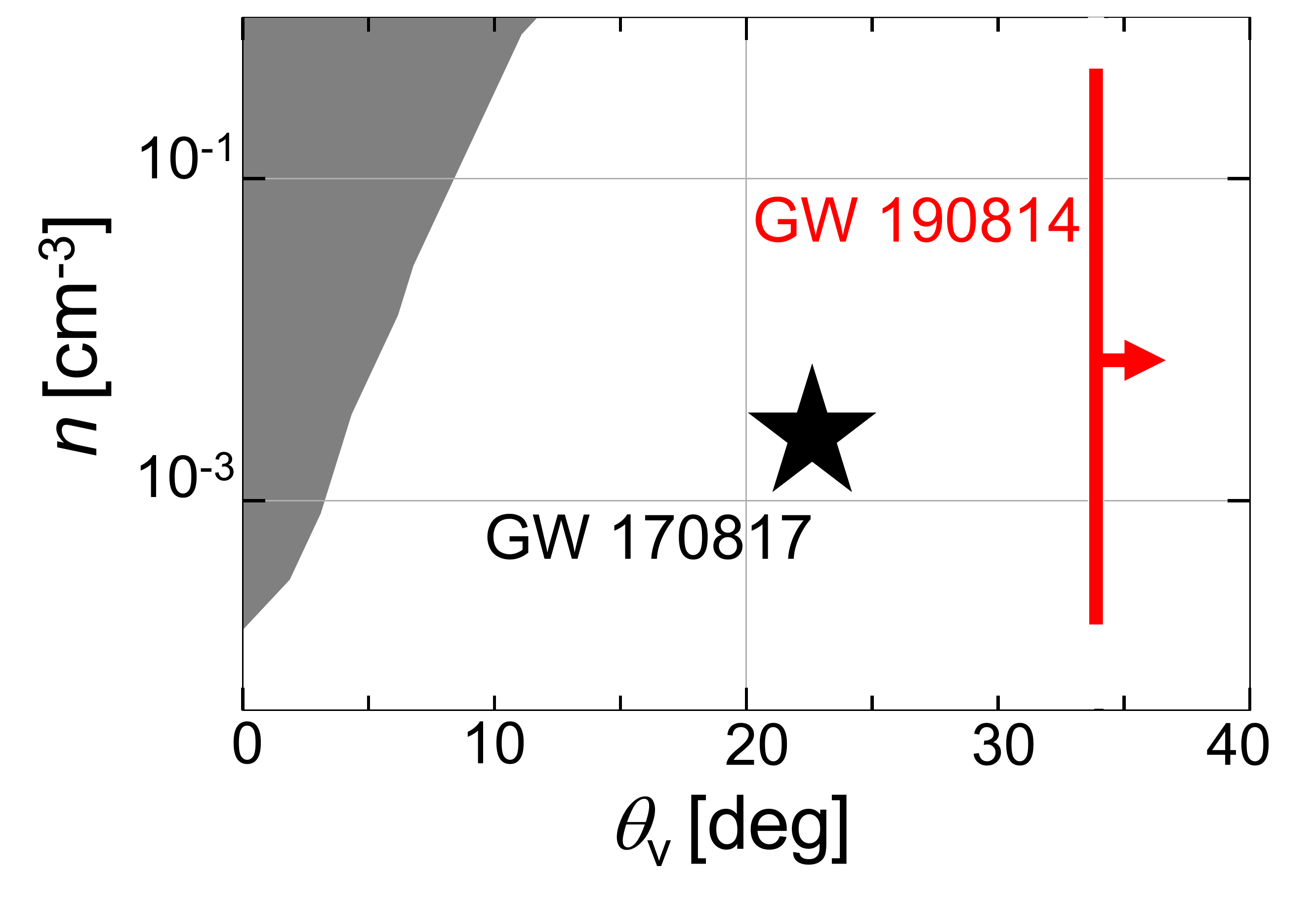}}}}
    \caption{ {\it Left panel}: Optical upper limits for GW190814 in comparison to observed short GRB afterglows. The light curves are scaled to a distance of 267 Mpc. Upper limits obtained from DDOTI and LDT are marked with filled symbols (green and red; respectively) and the upper limits from DECam and VST are marked with empty symbols (black and darkgreen; respectively). The afterglows that cannot be ruled out are coloured purple. 
    {\it Right panel}:  Optical upper limits for GW190814
    in comparison to off-axis afterglow light curves, calculated using the explosion properties of GW170817 \citep{Troja19,Ryan_2020} and viewing angles of $13\degree, 14\degree$ and $16 \degree$. 
    The inset shows the constraints on density as a function of the viewing angle. The region ruled out by the upper limits is shaded gray. The lower limit on the inclination for GW190814 is marked with a vertical red line. \label{fig:sGRBag}}
\end{figure*}

This factor may have been partially influenced by the behavior of AT2017gfo, which peaked at early times and then rapidly faded at optical wavelengths. The expectation of a weak signal probably discouraged the pursuit of additional spectroscopic observations. However, a wider range of kilonova peak times and decay rate is predicted by models (see Sect.~\ref{sec: KNprops}), and an improvement in late-time spectroscopic follow-up strategy could increase the chance of detecting kilonova signals differing from AT2017gfo. In the case of GW190814, a key factor may also have been the low probability of an EM signal, as calculated by the LVC preliminary analysis \citep{2019GCN.25333....1L}. 

It is worth noting that large aperture telescopes, such as the W. M. Keck Observatory \citep{2019GCN.25395....1D}, the Southern African Large Telescope \citep[SALT,][]{andreoni2019growth}, the Gran Telescopio de Canarias (GTC, this work) and the Very Large Telescope \citep[ VLT,][]{2020arXiv200201950A}, played a key role in securing the spectroscopic observations. 

\subsection{Constraints on afterglow emission and implications for the GRB jet}\label{jetag}

We use optical limits on the GW counterpart to constrain the presence of a relativistic jet component, as observed in short gamma-ray bursts (sGRBs).
In Figure~\ref{fig:sGRBag} (left panel), we compare the optical upper limits  to a sample of 21 sGRB afterglows  with known redshift,  rescaled to a distance of 267 Mpc \citep{2020MNRAS.492.5916W}.
The presence of a typical on-axis sGRB afterglow is disfavoured by the wide-field data: limits from DDOTI rule out 60\% of the light curves with a coverage of 88\% (corresponding to a 53\% probability). This is an improvement over the detection probability ($\sim$30\%) obtained by \citet{2020MNRAS.492.5916W}, as our limit is 1~mag deeper. The constraints from DECam and VST can exclude up to 80\% of the cases with a coverage of 92\% and 61\%, corresponding to a probability of 70\% and 49\%, respectively. 

This comparison is however based on detected sGRB afterglows, and may be biased toward the brighter end of the luminosity distribution. In order to assess our ability to constrain on-axis GRB explosions, we also ran a set of 10,000 simulations with input afterglow parameters representative of the broader sGRB population, including events without an observed optical counterpart. 
We adopt the standard framework of synchrotron emission from  shock-accelerated electrons with an energy distribution $N(E) \propto E^{-p}$ and $p$=2.2. Four parameters describe the afterglow behavior: the isotropic equivalent kinetic energy ($E_0$), the density of the external medium ($n$), the fraction of energy transferred to the electrons ($\epsilon_e$) and to the magnetic field ($\epsilon_B$). These parameters were randomly assigned assuming the observed distributions \citep{OConnor2020}, and simulated light curves for an on-axis top-hat jet were created using \texttt{afterglowpy}\footnote{https://github.com/geoffryan/afterglowpy} \citep{Ryan_2020}. The jet opening angle was fixed to a fiducial value of 5$^{\circ}$ \citep[e.g.,][]{Troja16,Jin18}.
Since the effects of collimation become apparent at $t$\,$\gtrsim$1~d,  this particular choice does not affect our conclusions, mostly driven by the early-time ($\approx$12~hrs) upper limits. 
Based on these simulations, a sizable fraction of on-axis afterglows can be confidently ruled out: we derive a 40\% probability from DECam constraints, a $\approx$30\% probability
from DDOTI and VST limits. 

Given the low rate of sGRBs in the local Universe \citep[e.g.][]{Dichiara2020}, the probability of intercepting an on-axis event is however very small. An off-axis explosion, that is a GRB jet not aligned to our line of sight, is a more likely counterpart of a GW source. Off-axis afterglows are much fainter than their on-axis counterparts, and could easily escape optical/nIR searches. 
For example, we consider the case of GW170817 and investigate whether a similar explosion could have been detected for GW190814. We use \texttt{afterglowpy} \citep{Ryan_2020} to simulate 2280
optical light curves with the same physical parameters derived for GW170817 and a range of viewing angles ($\theta_v$) and densities ($n$). For typical ISM densities $n$\,$\gtrsim$10$^{-4}$, the GW afterglow would have been detected if on-axis ($\theta_v$=0). However, the detection probability drastically decreases with increasing viewing angles, and becomes negligible for $\theta_v$\,$>$10~deg (see inset of \ref{fig:sGRBag}). 
Therefore, for the range of values derived from the GW data $\theta_v$ =45$^{+18}_{-11}$~deg \citep{Abbott_2020_GW190814bv}, 
any off-axis afterglow would have escaped detection.

\subsection{Constraints on kilonova ejecta properties} \label{sec: KNprops}

Optical and infrared observations constrain properties of a possible kilonova associated with GW190814. We compare upper limit observations to simulated kilonova light curves with varying input parameters corresponding to the distribution and properties of matter outside the remnant compact object. 
The amount of material ejected from an NSBH binary depends on the properties of the compact objects and, in particular, the BH mass and its spin.  
If the BH is not spinning, the total mass outside the remnant is roughly 0.1\,-\,0.2\,M$_\odot$ for a 4\,M$_\odot$ BH. This total ejecta mass decreases with increasing BH mass, falling to 0.01\,M$_\odot$ for a 7\,M$_\odot$ BH.  With co-rotating spins, this number increases to 0.2\,-\,0.3\,M$_\odot$ for a 4\,M$_\odot$ black hole \citep{2015PhRvD..92b4014K}.
Typically, the dynamical ejecta masses are a factor of 5 to 10 times lower than this total, with the remainder forming an accretion disk around the remnant compact object. Wind ejecta masses are roughly 10-30\% of the disk mass \citep{2019PhRvD.100b3008M}. However, only BHs with masses below 5\,M$_\odot$ and/or very large disk masses will produce sufficient disk outflows to sustain an observable kilonova~\citep{2020arXiv200514208F}. The final parameter estimates for GW190814 correspond to the merger of a 23\,M$_\odot$ black hole with a 2.6\,M$_\odot$ compact object \citep{Abbott_2020_GW190814bv} leaving little chance for matter outside the remnant BH, and thus significantly reduce the probability of producing an observable kilonova.

By comparing observational upper limits to kilonova light curve models, we can place independent constraints on the properties of both dynamical and wind ejecta components from this merger. 
Past studies of GW190814 have argued for a range of constraints.  Using a constant opacity model, \cite{andreoni2019growth} argued that the ejecta masses were less than 0.03-0.05\,M$_\odot$. Models using a more realistic opacity description and a two-component ejecta profile have argued that the ejecta masses above 0.1\,M$_\odot$ are typically ruled out and, depending upon the viewing angle, some ejecta masses as low as 0.04\,M$_\odot$ can be disfavored~\citep{2020ApJ...893..153K}.  As discussed below, our results, using a broader parameter range of morphologies and ejecta velocities, disfavour models with total ejecta masses above 0.1\,M$_\odot$ and, except for slow moving dynamical ejecta models, models with dynamical ejecta masses above 0.1\,M$_\odot$ are nearly all ruled out. Similarly, most models with wind mass exceeding 0.1\,M$_\odot$ and wind velocities above 0.15$c$ are also inconsistent with observations.

\begin{figure*}
\includegraphics[trim=15 5 35 15, clip, width=\columnwidth]{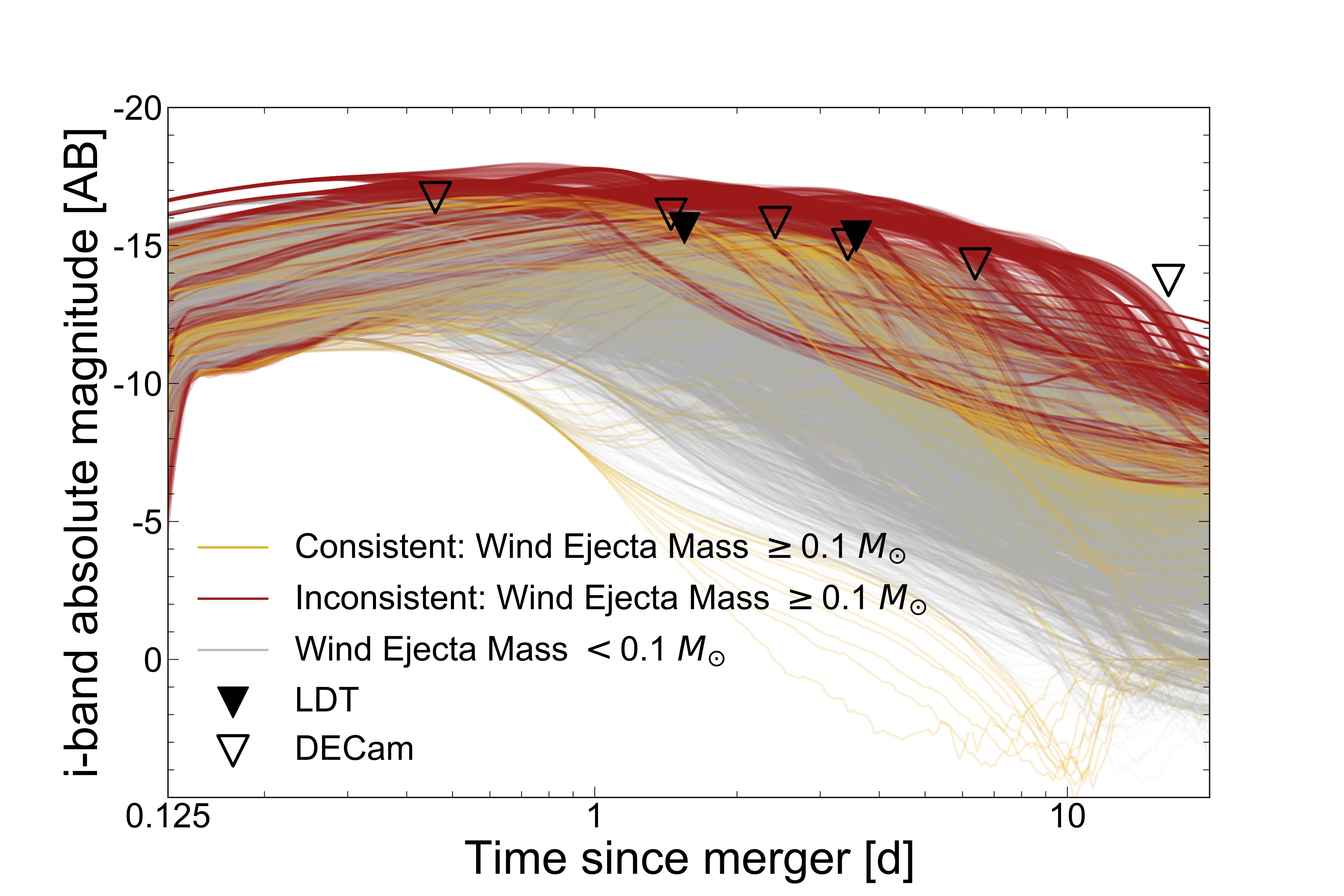}
\hspace{0.5cm}
\includegraphics[trim=15 5 35 15, clip, width=\columnwidth]{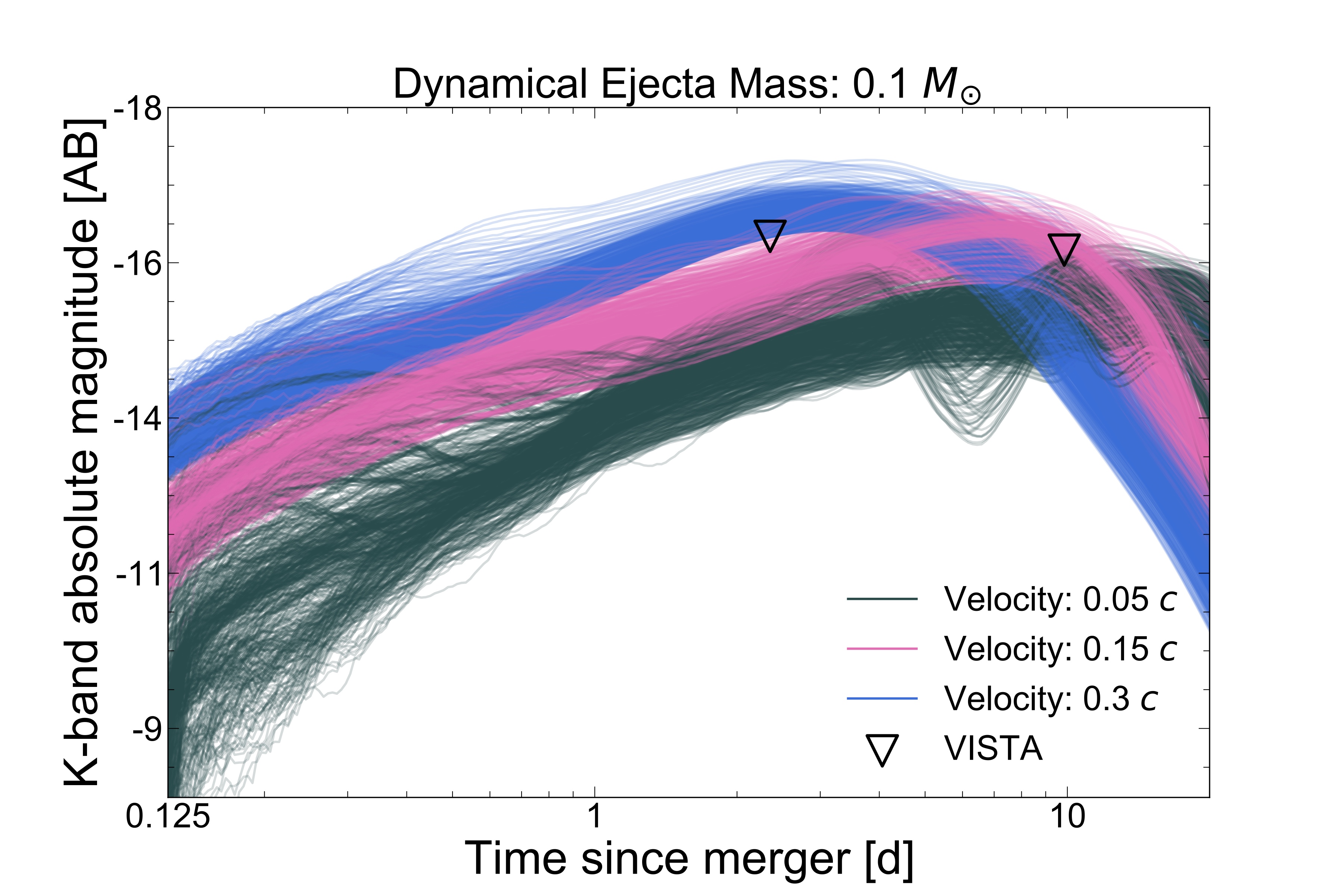}
\caption{Kilonova light curves from LANL simulation suite. For clarity, only 11 of the 54 viewing angles are presented, spanning every fifth angular bin of equal solid angle ranging from viewing angles of 0\degree to 180\degree. 
\textit{Left panel}: \textit{i}-band light curves for all kilonova parameters considered in the simulation space. Light curves are coloured by wind ejecta mass, with coloured (red or yellow) light curves containing a wind-mass of 0.1\,M$_\odot$ and gray light curves corresponding to kilonova with smaller wind ejecta masses. Wide-field upper limit constraints (open triangles) divide the light curves into those inconsistent with observational constraints (red) and those that remain feasible (yellow). Galaxy-targeted LDT upper limits (solid triangles) are not used to constrain parameters. 
\textit{Right panel}:  \textit{K}-band light curves with dynamical ejecta masses of 0.1\,M$_\odot$, with all other kilonova properties still allowed to vary. Light curves are differentiated by average dynamical ejecta velocity with the highest velocity in blue, intermediate velocity in pink, and lowest velocity in dark-green.}
\label{fig:lightcurves}
\end{figure*}

\subsubsection{Description of models}

For this study, we use a grid of two-component models from the LANL group (Wollaeger et al., in prep).  The two-components include a heavy r-process ejecta (a.k.a. dynamical ejecta) and either a high- ($Y_e=0.37$) or mid- ($Y_e=0.27$) latitude wind composition. The mid-latitude composition contains a trace abundance of lanthanides, while the high-latitude model produces no lanthanides. The morphology of these two components are set using the TS and TP profile shapes from a more extensive morphology study~\citep{2020arXiv200400102K}. These two morphologies assume a toroidal profile for dynamical ejecta and either a spherical or peanut-shaped profile for the wind. The light curves from these models use the \texttt{SuperNu}~\citep{2014ApJS..214...28W} code that has now been run in a wide range of supernova and kilonova studies~\citep{2018MNRAS.478.3298W,2019arXiv190413298E,2019ApJ...880...22W,2020arXiv200400102K}. \texttt{SuperNu} is a multi-dimensional, multi-group Monte Carlo transport scheme, which produces light curves for a broad range of viewing angles. In addition, we employ the WinNet nucleosythesis network~\citep{2012ApJ...750L..22W} to simulate heating from radioactive decay of our prescribed abundances. The opacities use the latest LANL opacity database: a full set of lanthanide opacities from \cite{2020MNRAS.493.4143F} with uranium acting as a proxy for the full set of actinides. 
The grid of models includes two morphologies and two wind compositions in addition to a range of dynamical ejecta and wind masses (0.001, 0.003, 0.01, 0.03, 0.1 $M_\odot$). The grid also includes ejecta velocities of 0.05$c$, 0.15$c$, and 0.3$c$, corresponding to peak ejecta velocities of 0.1$c$, 0.3$c$, and 0.6$c$. The grid varies all six parameters independently, creating 900 different explosion models. Light curves depend on the viewing angle due to non-spherical morphologies; thus, we consider 54 different viewing angles for each model. The 54 polar viewing angles range from on-axis (0\degree) through edge-on (90\degree) and back to on-axis (180\degree), subtending an equal solid angle in each angular bin. Light curves are not binned in the azimuthal direction, due to the axisymmetric nature of the simulations. Including the angular dependence, we have 48,600 different sets of time-dependent kilonova spectra in our simulation database to compare to the observational constraints.

\subsubsection{Model comparison to data}
This work expands upon past studies of GW190814 by both including the full set of observational limits and utilizing a broad grid of two-component models with realistic opacities. Our state-of-the-art grid produces a much more diverse set of light curves than past studies of these events.  In this section, we assume negligible contamination from any GRB afterglows and that the possible kilonova dominates the observed \textit{i}, \textit{r}, \textit{J} and \textit{K}-band emission. 
As discussed in the previous section, this assumption is well justified by the lack of any on-axis GRB as well as the expected faintness of an off-axis afterglow component. 

Wide-field upper limits place the most compelling constraints on the data, and will be the focus of our kilonova parameter constraints. We direct our analysis to DECam upper limits in the \textit{i}-band, VST upper limits in the \textit{r}-band, and the VISTA upper limits in \textit{K}-band. All upper limits are scaled to absolute magnitudes assuming a median luminosity distance of 267 Mpc.

Figure~\ref{fig:lightcurves} shows a subset of simulated \textit{i} (left panel) and \textit{K}-band (right panel) light curves (11 of the 54 viewing angles) compared to observational constraints.  These light curves follow many of the same trends expected in transients.
For example, models with faster ejecta velocities expand more quickly, causing earlier rise and fall times as well as brighter peak emission.  The early-time \textit{i}-band emission is dominated by the wind ejecta ("blue" component) and the late time \textit{K}-band emission is dominated by the dynamical ejecta ("red" component).  In an ideal scenario, observations would reveal a simple correspondence between \textit{i}-band luminosity and the wind ejecta mass/velocity and, similarly, a relation between \textit{K}-band luminosity and dynamical ejecta mass/velocity. However, additional properties affect the emission and further obscure this relationship. These properties include distribution of ejecta (e.g., morphology), lanthanide curtaining where the dynamical ejecta obscures the wind material and alters the early-time emission, and variations in the abundances. In general, models with more ejecta mass are brighter and are thus ruled out by the upper limit constraints.

\begin{figure}
\begin{center}
\includegraphics[trim=15 40 0 90, clip, width=\columnwidth]{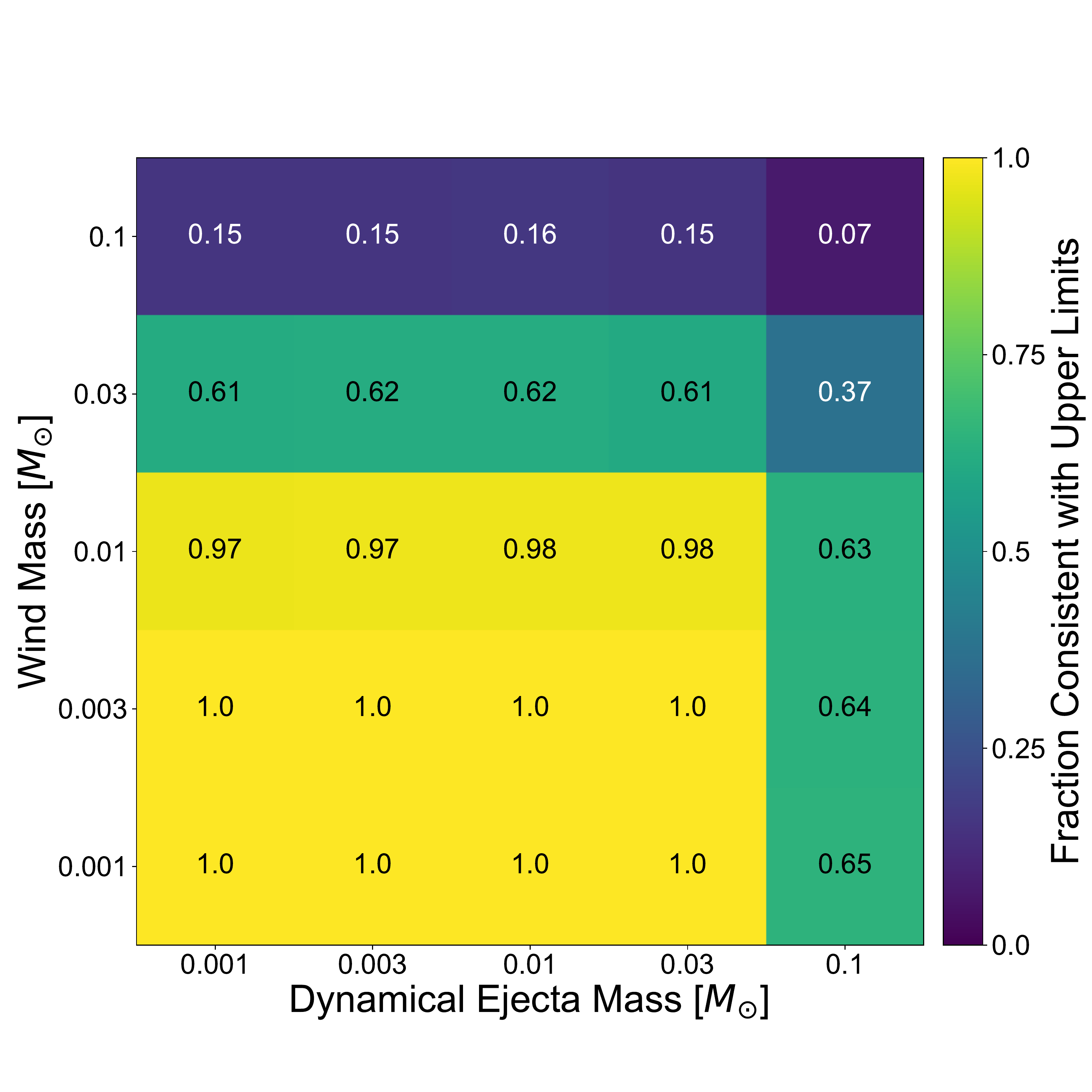}
\caption{Fraction of simulated kilonovae consistent with wide-field upper limit constraints. We separate kilonova simulations by their input dynamical ejecta masses and wind masses. Each mass combination is represented by a colour and decimal fraction indicating the percentage of simulations for that given set of parameters that remain consistent with upper limits. We evaluate 1944 plausible kilonova simulations for each mass combination, with varying viewing angle, wind composition model, wind ejecta morphology, and velocities.
}
\label{fig:grid_masses}
\end{center}
\end{figure}

The \textit{i}-band light curves (Figure~\ref{fig:lightcurves}, left panel) are dominated by the wind ejecta. 
The colour-coding is based on the wind ejecta mass:  gray models correspond to simulations with ejecta masses below 0.1\,M$_\odot$, 
coloured models to simulations with ejecta masses above 0.1\,M$_\odot$. 
The high-mass ejecta models are further delineated by whether they are ruled out by the observed upper limits: red models have luminosities that exceed at least one upper limit (ruled out by the data), yellow models lie below all the wide-field data (consistent with the data).  

The dynamical ejecta plays a more important role in shaping the \textit{K}-band light curves (Figure~\ref{fig:lightcurves}, right panel).  The fast-velocity (average velocity of 0.3$c$), 0.1\,M$_\odot$ dynamical mass models are nearly all ruled out by the VISTA upper limit at $\sim$2.35\,d. Roughly 35\% of all intermediate velocity (average velocity of 0.15$c$) models with 0.1\,M$_\odot$ dynamical mass are inconsistent with the constraints
at $\sim$9.85\,d. Due to their later peak time, slightly more low-velocity models remain plausible.

As many factors contribute to the light curve morphology, we cannot prescribe a one-to-one correspondence between upper limits and a specific component of the ejecta.  Figure~\ref{fig:grid_masses} shows the fraction of models consistent with the observed upper limits.  Less than 15\% of our massive (0.1\,M$_\odot$) wind ejecta models are consistent with the data and only 7\% of these models with high dynamical ejecta masses (0.1\,M$_\odot$) lie below these limits.  On the other extreme, all models with wind ejecta masses below 0.01\,M$_\odot$ and dynamical ejecta masses below 0.1\,M$_\odot$ are consistent with the data.  Given the estimates of the BH mass \citep{Abbott_2020_GW190814bv}, the constraints on the ejecta masses are consistent with the expectations from merger simulations.  

\begin{figure*}
\centering
\includegraphics[width=0.8\columnwidth]{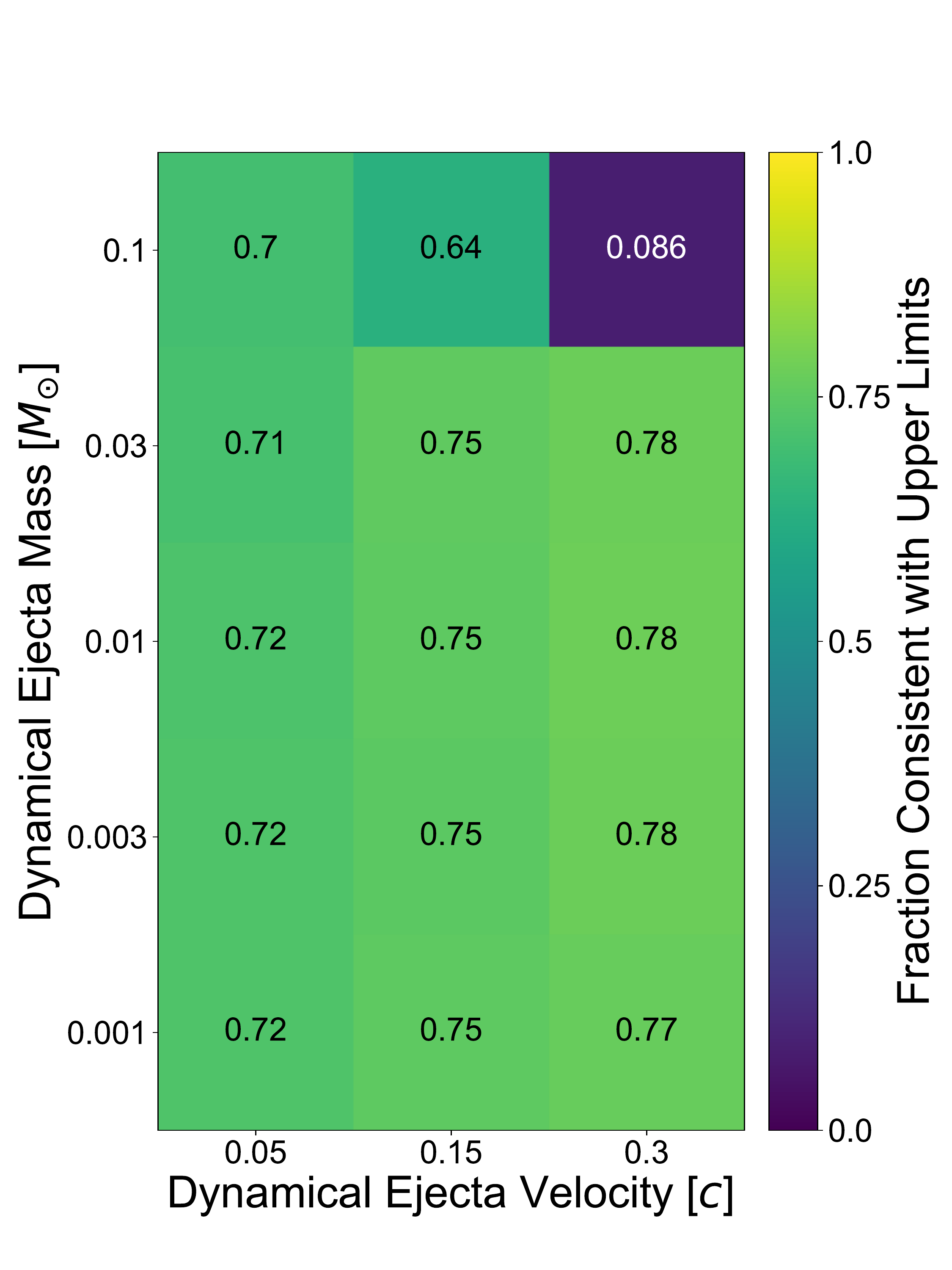}
\hspace{0.4cm}
\includegraphics[width=0.8\columnwidth]{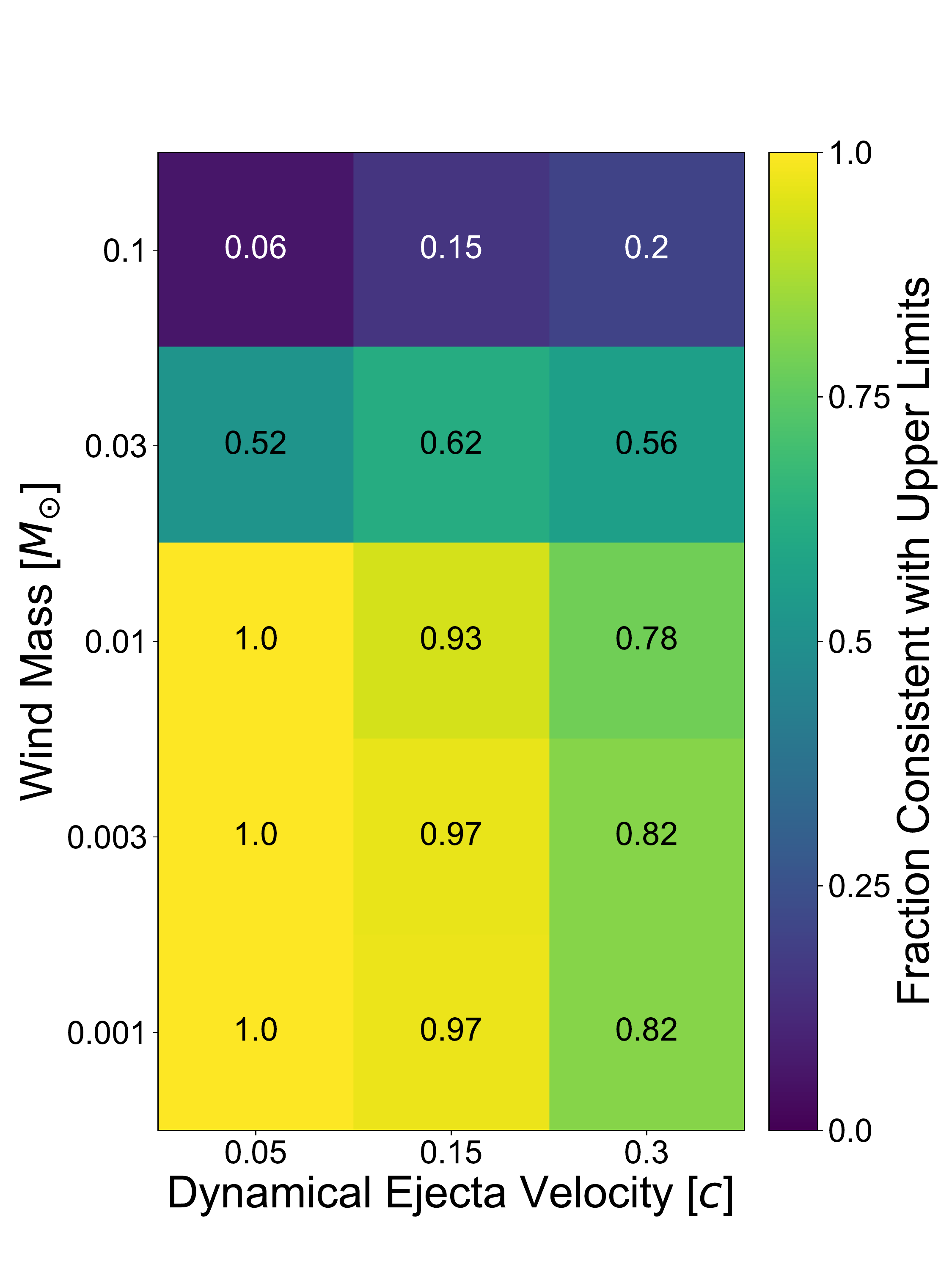}
\includegraphics[width=0.8\columnwidth]{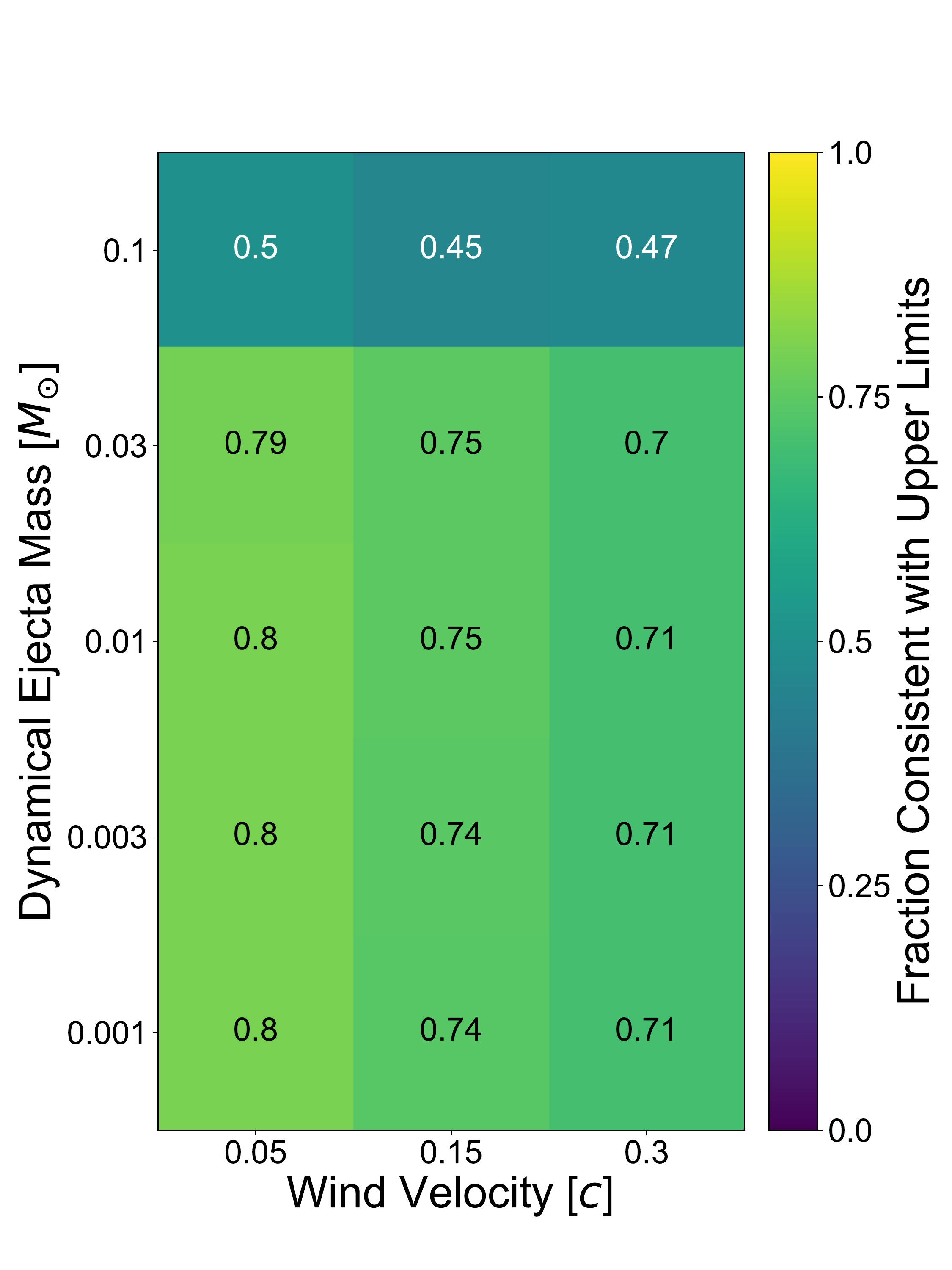}
\hspace{0.4cm}
\includegraphics[width=0.8\columnwidth]{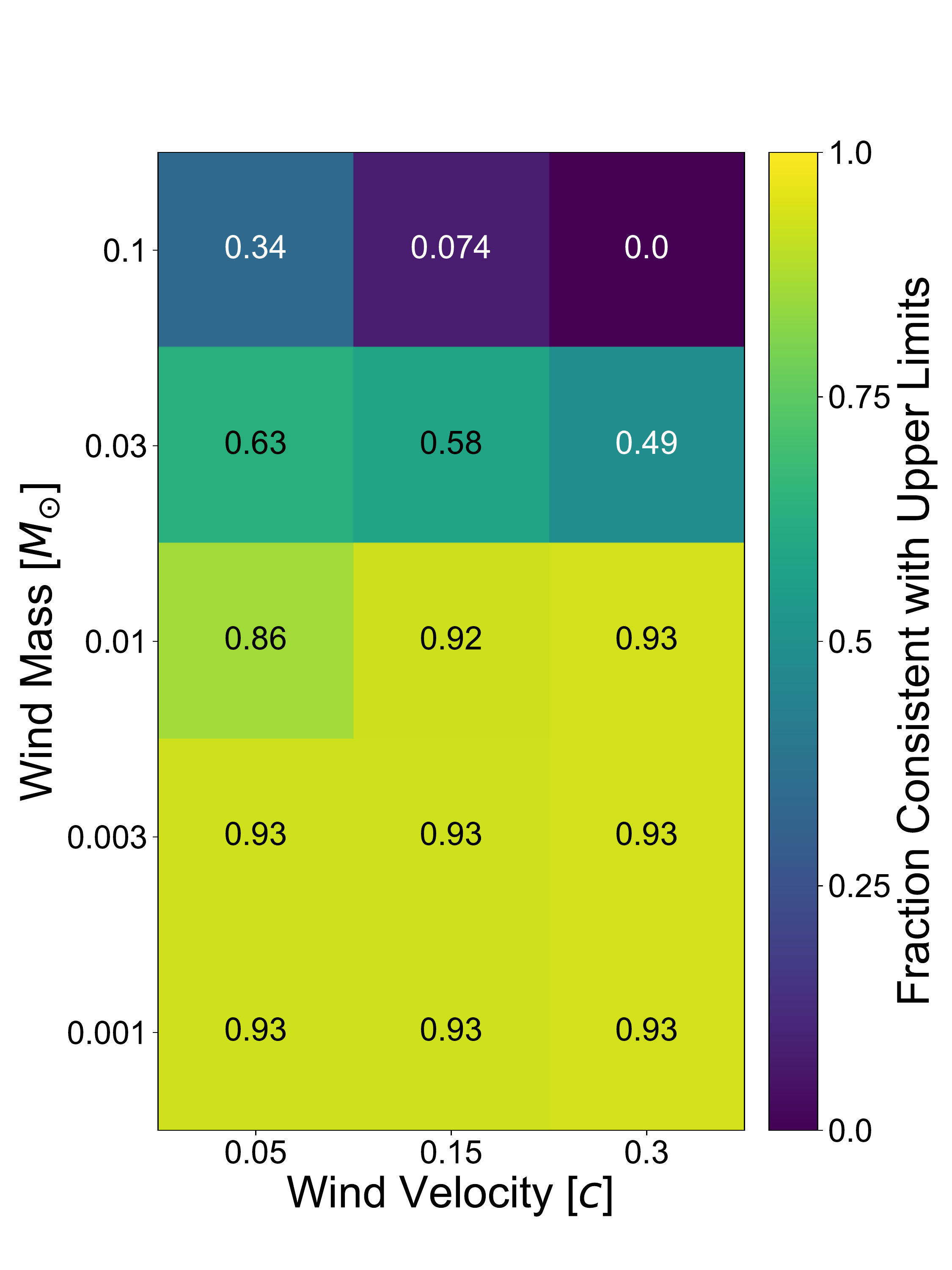}
\caption{Fraction of simulated kilonovae consistent with wide-field upper limit constraints. Constrains are displayed jointly for each combination of mass and velocity parameters. Each parameter combination is represented by 3240 plausible kilonova simulations, with varying viewing angle, wind composition model, wind ejecta morphology, and masses and velocities not directly shown in each subfigure. The following paramter combinations are displayed: \textit{Top left panel:} dynamical eject mass and velocity, \textit{Top right panel:} wind mass and dynamical ejecta velocity, \textit{Bottom left panel:} dynamical ejecta mass and wind velocity, \textit{Bottom right panel:} wind mass and velocity.}
\label{fig:grid_withvelocities}
\end{figure*}

Figure~\ref{fig:grid_withvelocities} shows the fraction of consistent models studying different parameters. In these images, we reiterate many of the results seen in our light curve models. The top left panel demonstrates that the majority of fast (early-peaking), high-mass dynamical ejecta models are ruled out, as anticipated from the right panel of Figure~\ref{fig:lightcurves}.  However, fast dynamical ejecta can obscure the signal from wind ejecta,
and 34\% of the 0.1\,M$_\odot$ slow-moving wind ejecta models are consistent with the observations versus only 6\% for the slow dynamical ejecta with comparable wind ejecta mass.  
Similarly, fast wind ejecta models both extend beyond the dynamical ejecta (this emission is not blocked) and peak brighter and earlier (ruled out by early observations).  All fast-moving wind models with 0.1\,M$_\odot$ wind mass ejecta are ruled out by the current constraints.

Of the 12 candidates with an unknown classification that have redshifts consistent with the GW distance scale (See Section~\ref{sec:Opticalfollowup}), 10 have reported $i$-band detections and we compared them to our grid of simulated kilonova light curves. Two sources, AT2019tiw and AT2019tij, are inconsistent with all simulated kilonovae, remaining 2 magnitudes brighter than any plausible light curve. Some detections correspond to high wind ejecta masses ($\geq$ 0.1M$_\odot$) and low wind velocities ($\leq$ 0.15$c$). None of these candidates provide strong constraints on either dynamical ejecta mass or velocity.

We do not consider here the effects of the total extinction along the line of sight. However, we do not expect it to substantially change our conclusions since extinction is less important in the infrared band. Furthermore, short GRBs and GW170817 were found in sites of low extinction \citep[e.g. ][]{Levan2017}.

\section{Conclusions}\label{sec:con}

We have presented here our search for possible optical/nIR emission from GW190814. 
Our wide-field DDOTI observations covering 88\% of the probability area did not find a potential counterpart up to a limiting magnitude $w_{max}$ $\approx$ 19 AB mag. 
Targeted observations of nearby galaxies were carried out using the Lowell Discovery Telescope, and did not identify any candidate counterpart down to $i = 22.9$ AB mag. 
Additionally, our RATIR and GTC observations focused on classification of candidates through multi-colour photometry and spectroscopy, respectively. We could not find any association to a possible kilonova for all the candidates that were covered by these observations.

A total of 85 optical transients, with brightness ranging between 18 and 24 AB mag, were identified by other searches as possible counterparts of GW1901814. We find that about 75\%  of these can be ruled out, while the remaining 21 objects are left unclassified. We find the follow-up observations to be very thorough within the first four days post-merger, with $\sim 67 \%$ of the candidates announced in this period having a spectroscopic classification.  A successful source identification was less likely at later times. 
These findings highlight that, even for well-localized events 
of high-interest such as GW190814, the identification efficiency of the follow-up campaign is lower than its detection efficiency. In addition to the sensitivity of the observing facilities, other factors, such as variable observing conditions, limited allocated time, or delays in the source classifications, further reduce the chances to find and identify the GW counterpart.

We used our observations in conjunction with the community-wide follow-up observations to place constraints on the GRB afterglow emission and the kilonova parameter space for this event. 
On-axis afterglows are strongly disfavored, which is in line with the non-detection of gamma-ray emission. Off-axis afterglow light curves are instead too faint to be meaningfully constrained.
In particular, for an energetic explosion similar to GW170817, 
any viewing angle above 10~deg would be consistent with the observations.

Based on our extensive kilonova simulations grid, we could constrain a wide range of ejecta masses and velocities. 
We find that models with high wind masses (0.1 M$_{\odot}$) and high dynamical ejecta masses ($\gtrsim$ 0.1 M$_{\odot}$) are disfavoured by the optical upper limits. Additionally, nIR upper limits disfavour fast moving ($\geq$ 0.3$c$) dynamical ejecta, assuming the dynamical ejecta mass to be 0.1 M$_{\odot}$.

Thanks to the large set of kilonova simulations, we find that a broader range of ejecta masses can be consistent with the data than past studies. For example we can not rule out all of our models with 0.1\,M$_\odot$ ejecta (although we rule out most of these high-mass models).  But the observations do rule out most of the wind ejecta (high electron fraction material from the disk) models above 0.1\,M$_\odot$ and the fast-moving, high-mass dynamical ejecta (low electron fraction).  These constraints are consistent with the latest models of ejecta masses from NSBH mergers~\citep{2020arXiv200514208F}.

The recently published parameter values for this merger, a 23~M$_{\odot}$ BH merging with a 2.6 M$_{\odot}$ compact object, have interesting implications for the possible EM counterparts, supporting scenarios that encompass little or negligible ejecta. The high mass ratio suggests that there is a low chance of remnant matter outside the final object as the more massive BH will likely directly absorb the secondary component without its disruption. Furthermore, the nature of the secondary component is unclear from GW observations, and a low mass BH cannot be ruled out.

With the upcoming increase in sensitivity and addition of new detectors to the global GW network, 
we can expect future GW detections with smaller localization regions and at even farther distances \citep{2018LRR....21....3A,2019arXiv190912961P}. 
The case of GW190814 shows that, despite its good sky localization, small to medium aperture ground-based detectors are challenged at 
distance scales $\gtrsim$200~Mpc, and can only probe the brightest end of the luminosity distribution, corresponding to nearly-on axis GRB afterglows and high-mass kilonova ejecta. 
In the case of GW190814, the inclination angle of $\approx$\,45~deg and the high mass ratio of the binary components derived from the GW signal are not favorable to the detection of an EM counterpart, consistent with the lack of any suitable candidate from an extensive follow-up campaign.  Information on the merging binary properties, such as 
its inclination and mass ratio, would therefore be a critical input for the observing community in order to optimize the use of observational resources as well as the subsequent effort of data analysis and source classification.

\section*{Acknowledgements}

ALT, RSR and LP acknowledge support from the European Union’s Horizon 2020 Programme under the AHEAD2020 project (grant agreement n. 871158) and by ASI (Italian Space Agency) through the Contract no. 2019-27-HH.0 and from MIUR (PRIN 2017 grant 20179ZF5KS). SD and ET acknowledge support for this work under NASA grant 80NSSC18K0429. 
JBG acknowledges the support of the Viera y Clavijo program funded by ACIISI and ULL. GB acknowledges financial support under the INTEGRAL ASI-INAF agreement 2019-35-HH.0

We thank Charlie Hoy for his help in accessing and loading the finalized GW190814 skymap.

We thank the staff of the Observatorio Astron\'omico Nacional. Some of the data presented in this paper were acquired with the DDOTI instrument of the Observatorio Astron\'omico Nacional on the Sierra de San Pedro M\'artir. DDOTI is funded by CONACyT (LN 260369, LN 271117, and 277901), NASA Goddard space Flight center, the University of Maryland (NNX17AK54G), and the Universidad Nacional Aut\'onoma de M\'exico (CIC and DGAPA/PAPIIT IT102715, IG100414, AG100317, and IN109418) and is operated and maintained by the Observatorio Astron\'omico Nacional and the Instituto de Astronom\'ia of the Universidad Nacional Aut\'onoma de M\'exico.
We acknowledge the contribution of Neil Gehrels to the development of DDOTI.

Some of the data used in this paper were acquired with the RATIR instrument, funded by the University of California and NASA Goddard Space Flight Center, and the 1.5-meter Harold L.\ Johnson telescope at the Observatorio Astron\'omico Nacional on the Sierra de San Pedro M\'artir, operated and maintained by the Observatorio Astron\'omico Nacional and the Instituto de Astronom{\'\i}a of the Universidad Nacional Aut\'onoma de M\'exico. We acknowledge the contribution of Leonid Georgiev and Neil Gehrels to the development of RATIR.

The spectroscopic data presented in this work were reduced using standard routines of PyRAF. PyRAF is a product of the Space Telescope Science Institute, which is operated by AURA for NASA. This research made use of ccdproc, an Astropy package for image reduction \citep{matt_craig_2017_1069648}. This work made use of the data products generated by the NYU SN group, and released under DOI:10.5281/zenodo.58766, available at \url{https://github.com/nyusngroup/SESNtemple/}.

\section*{Data Availability}
The data underlying this article will be shared on reasonable request to the corresponding author.

%%%%%%%%%%%%%%%%%%%% REFERENCES %%%%%%%%%%%%%%%%%%

% The best way to enter references is to use BibTeX:

\bibliographystyle{mnras}
\bibliography{References} % if your bibtex file is called example.bib

% Alternatively you could enter them by hand, like this:
% This method is tedious and prone to error if you have lots of references
%\begin{thebibliography}{99}
%\bibitem[\protect\citeauthoryear{Author}{2012}]{Author2012}
%Author A.~N., 2013, Journal of Improbable Astronomy, 1, 1
%\bibitem[\protect\citeauthoryear{Others}{2013}]{Others2013}
%Others S., 2012, Journal of Interesting Stuff, 17, 198
%\end{thebibliography}

%%%%%%%%%%%%%%%%%%%%%%%%%%%%%%%%%%%%%%%%%%%%%%%%%%

%%%%%%%%%%%%%%%%% APPENDICES %%%%%%%%%%%%%%%%%%%%%

%\appendix

%\section{Some extra material}

%If you want to present additional material which would interrupt the flow of the main paper,
%it can be placed in an Appendix which appears after the list of references.

%%%%%%%%%%%%%%%%%%%%%%%%%%%%%%%%%%%%%%%%%%%%%%%%%%

% Don't change these lines
\bsp	% typesetting comment
\label{lastpage}
\end{document}